\documentclass[%
reprint,
 amsmath,amssymb,
 aps,
]{revtex4-2}

\usepackage{graphicx}
\usepackage{dcolumn}
\usepackage{bm}
\usepackage{hyperref}
\usepackage{epstopdf} 
\usepackage{float}
\usepackage{comment}
\usepackage{url}
\usepackage{tikz}
\usepackage{subfigure}

\begin{document}

\preprint{APS/123-QED}

\title{Yielding behaviour of glasses under shear deformation at constant pressure}

\author{Krishna K Tiwari}
\author{Srikanth Sastry}%
 \email{sastry@jncasr.ac.in}
\affiliation{%
Jawaharlal Nehru Centre for Advanced Scientific Research, Jakkur Campus, Bengaluru, 560064, India 
}%

\date{\today}

\begin{abstract}
Computer simulations of yielding of glasses under shear have typically been performed under constant volume, strain controlled protocols. However, volumetric effects, such as the dilatancy associated with plastic rearrangements, and the observed reduction of density in shear bands, make it interesting to consider constant pressure shear protocols. We present a computational investigation on the nature of yielding of glasses under constant-pressure conditions, for different pressures. For uniform shear, the stress-strain curves at different pressures differ only by the stress scale. We find stable shear bands under cyclic shear whose steady-state width increases with an increase in external pressure, with density within shear bands being lower compared to the average values reached. Cyclically sheared well annealed glasses yield with a discontinuous dilation at the yield point, whereas the poorly annealed glasses undergo compaction before yielding accompanied by dilation. The external pressure influences the quantitative mechanical response of the glasses, but the qualitative behaviour is similar at different pressures, and remains the same as that of yielding at the constant-volume strain-controlled conditions. We discuss directions along with further investigations may be pursued, based on the results presented. 
\end{abstract}

\maketitle



\section{INTRODUCTION}



Amorphous solids form a diverse range of materials, including molecular and polymeric glasses, colloids, foams, and granular packings, all of which lack the long-range periodic order that characterizes crystals. Despite their diversity, these materials share several qualitative aspects of their mechanical response to external stresses or deformation; and the differences in their response has in recent times been attempted to be understood in terms of differences in their preparation history as an important factor \cite{BonnRevModPhys2017, NicolasRevModPhys2018,Parmar2019b}. Key mechanical properties, from the perspective of amorphous solids being structural materials, concern the nature of plasticity, yielding, and how they may lead to failure. Extensive investigations involving numerical simulations \cite{maloneypre06, karmakarpre10, jaiswalprl16, fiocco2013, priezjev2013, regev2015reversibility, leishangthemNAT2017, Jin, ozawapnas2018, ParmarPRX2019, barbotpre20}, theoretical  \cite{Dasgupta2012, LinPNAS14, parisipnas17, urbani2017b, BudrikisNATCOM2017, popovic2018a, barlow2020, liu2020oscillatory, sastryPRL20, khirallahPRL20, ParleyPhysFluid2020, MunganPRL2021, sarkar2025} and experimental investigations \cite{SunPRL10, AntonagliaPRL14, Lu2018, Zeng2018} have sought to characterize the nature of the yielding transition, focusing typically on the application of shear deformation. Computer simulations have typically employed strain controlled deformation at constant global volume. However, it has long been well appreciated that volumetric effects are an important component of the mechanical response of amorphous solids \cite{Reynolds1885,Lemaitre2002,Ren2013,Wang2016,Jiang2017,Lu2018,Zeng2018,Sun2019,Wang2016,Wang2021,Babu2021,Moriel2024}. Although extensively discussed in the context of  granular matter \cite{Reynolds1885,Ren2013,Babu2021}, dilatational effects associated with shear induced plasticity have been well appreciated in the context of glasses, particularly metallic glasses \cite{Lemaitre2002,Jiang2017,Wang2016,Wang2021}, and the importance of investigating dilatational plasticity in a wider context has also been emphasized \cite{Moriel2024}. In the context of cyclic shear yielding, it has been observed in simulations \cite{parmar2019} that the density of yielded glasses within the shear band, wherein strain is localised, is lower, which is another indication of the strong correlation between plasticity and dilation. However, since typical simulations are performed at constant global volume, the simulation protocols impose a constraint on the heterogeneous density behaviour observed. It is thus of interest to consider the yielding behaviour under constant pressure conditions. We perform athermal quasistatic shear (AQS) simulations under constant pressure conditions, investigating both uniform and cyclic shear deformation. Our results reveal that although quantitative details change under the constant pressure conditions, the qualitative behaviour associated with uniform and cyclic shear yielding are not affected. In particular, shear bands formed under cyclic shear remain stable and reach widths that remain stable after several cycles of shear. 

In the following, we first details of the investigated model and methods in Sec. II. Sec. III contains results, and we close with a summary and discussion in Sec. IV.

\section{MODEL AND METHODS}
We simulate the 80:20 Kob-Andersen binary Lennard-Jones mixture of particles(4000 and 64000) at applied reduced pressures of $2$, $10^{-3}$ and $2$. Lennard-Jones interaction with quadratic corrections to the potential energy and force is used, such that they smoothly go to zero at the interaction cutoff distance of $2.5$ times the interparticle diameter.

\begin{eqnarray}
U_{\alpha \beta} (r) &=& 4\epsilon_{\alpha \beta} \left[ \left( \frac{\sigma_{\alpha \beta}}{r} \right)^{12} - \left( \frac{\sigma_{\alpha \beta}}{r} \right)^{6}      \right] \nonumber \\
& & + 4\epsilon_{\alpha \beta} \left[ c_{0\alpha \beta} + c_{2\alpha \beta} \left( \frac{r}{\sigma_{\alpha \beta} } \right)^2   \right], ~~~~~r_{\alpha \beta} < r_{c\alpha \beta} \nonumber \\
&=& 0,~~~~~ \text{otherwise}.
\end{eqnarray}

Here, indices $\alpha, \beta $ are indices over the particle types, $c_{0\alpha \beta}$ and $c_{2\alpha \beta}$ are coefficients for quadratic corrections. The interactions parameters are defined in terms of $``A"$ particle parameters, as $\epsilon_{AB}/\epsilon_{AA} = 1.5$, $\epsilon_{BB}/\epsilon_{AA} = 0.5$, $\sigma_{AB}/\sigma_{AA} = 0.80$, $\sigma_{BB}/\sigma_{AA} = 0.88$. Length and energy are expressed in units of $\sigma_{AA}$ and $\epsilon_{AA}$, respectively. Poorly annealed samples are generated by equilibrating random configurations of particles at temperature T = 0.66 using constant volume, temperature molecular dynamics simulations using the Nos\'e-Hoover thermostat at reduced density $\rho=1.2$. The equilibration step is followed by energy minimization using the conjugate gradient method, leading to average energy minimum (inherent structure) energies $e_{IS} = -6.92$. Well annealed samples at $e_{IS}=-7.05$ are obtained by employing mechanical annealing using the method reported in \cite{das_parmar_22} (see \cite{BhaumikPNAS2021} for further details). Initial configurations at the different pressures are then obtained by minimizing the enthalpy  ($H=U+PV$) at the desired pressure.  
Although the energies change as a result of constant pressure minimization, we refer to poorly annealed and well annealed glasses with labels $e_{IS} = -6.92$ and $e_{IS} = -7.05$ for simplicity. Further, we note that while there may be other protocols for preparing glasses under constant pressure conditions, we adopt the procedure outlines above since our goal is to investigate the role of constant pressure deformation rather than exploring their preparation under different conditions. 

The shear deformations are carried out using the athermal quasistatic shear (AQS) simulations. A small strain increment of $d\gamma_{xz}=2\times 10^{-4}$ along the xz plane ($x^\prime = x+z d\gamma_{xz}$, $y^\prime=y$, $z^\prime=z$) is applied, followed by an energy minimization step for constant volume AQS simulations. In constant pressure AQS, the energy minimization is replaced by an enthalpy ($H=U+PV$) minimization at the imposed external pressure. In both cases, the minimization step terminates when the force or the energy tolerance of $10^{-13}$ is reached. For cyclic shear deformation, the strain of the system varies as $0 \rightarrow \gamma_{max} \rightarrow 0 \rightarrow -\gamma_{max} \rightarrow 0$, whereas for uniform shear deformation, the strain of the system is increased in shear steps of $d\gamma_{xz}$ up to the target strain. All the numerical simulations, including molecular dynamics, cyclic and uniform shear, and the energy/enthalpy minimization, are performed using LAMMPS \cite{LAMMPS}.


The reported data for uniform shear is averaged over $40$ well annealed samples of $4000$ particles and inherent energy of $-7.05$. The data for $64000$ particles is averaged over $5$ samples. For comparison, the results from constant volume shear simulations at density $\rho = 1.2$ are shown in the figures. In cyclic shear simulations, the steady state values  of observables are obtained by fitting to the data points a stretched exponential function.


\section{RESULTS}

\begin{figure}
    \centering
        \begin{tikzpicture}
            \node[anchor=south west, inner sep=0] (image) at (0,0) {\includegraphics[width=0.49\linewidth]{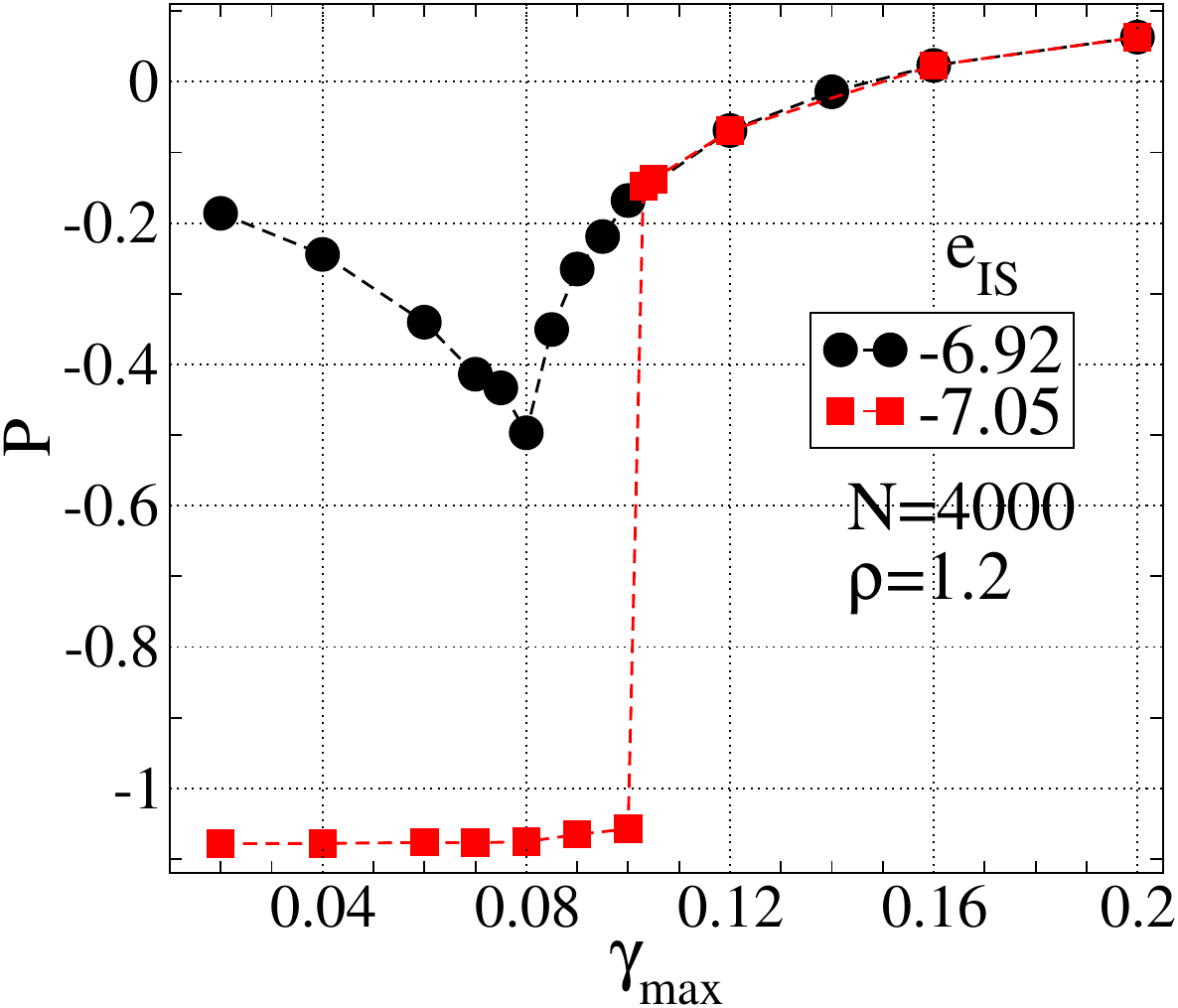}};
            \node[anchor=north west, font=\large\bfseries] at (image.north west) [xshift=0mm, yshift=7mm] {a)};
        \end{tikzpicture}
    \hfill
        \begin{tikzpicture}
            \node[anchor=south west, inner sep=0] (image) at (0,0) {\includegraphics[width=0.49\linewidth]{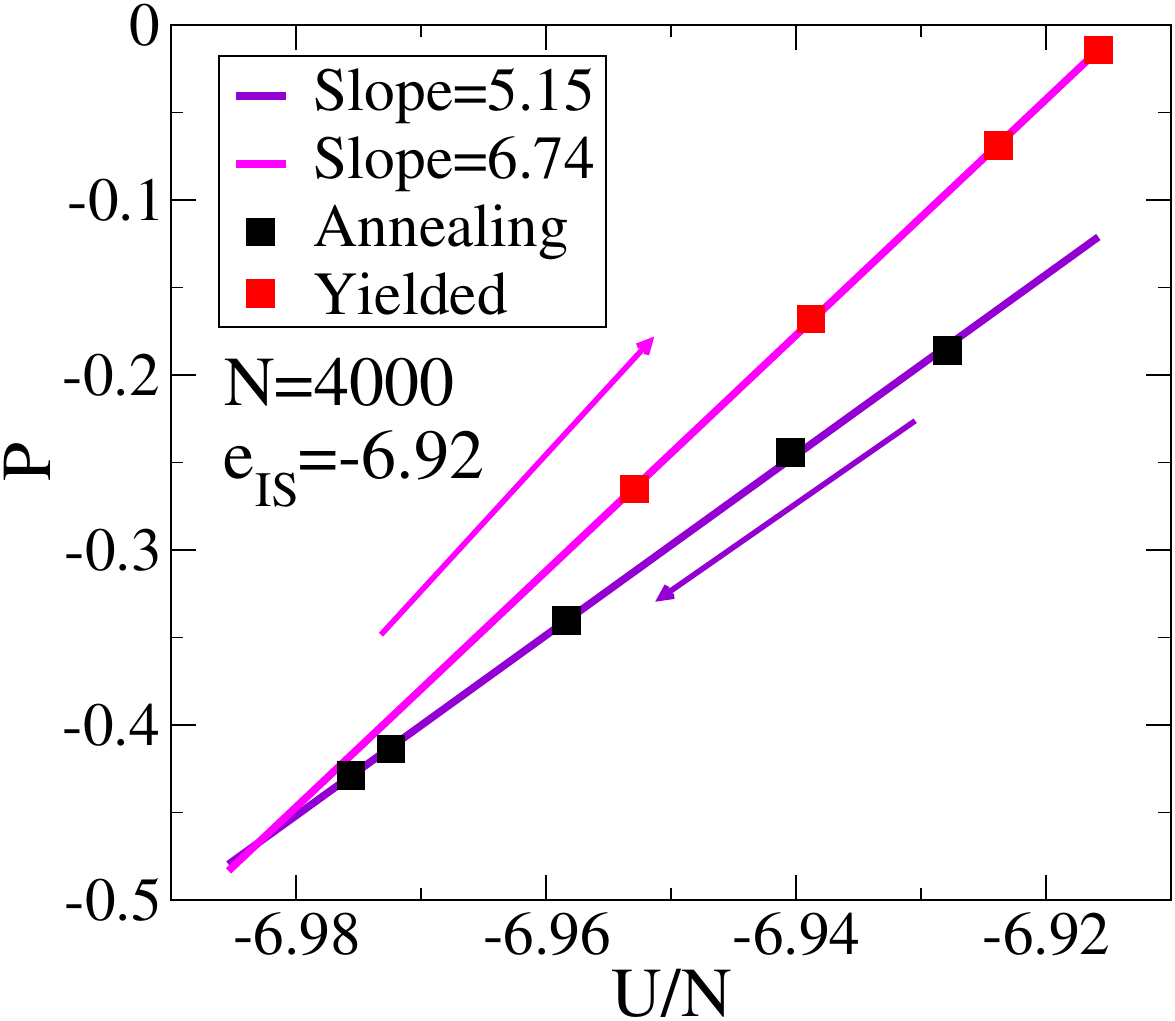}};
            \node[anchor=north west, font=\large\bfseries] at (image.north west) [xshift=0mm, yshift=7mm] {b)};
        \end{tikzpicture}
    \caption{\textbf{Variation of pressure for constant volume shear:} a) Pressure decreases for a poorly annealed sample till the yield shear amplitude is reached. For a well annealed sample, it varies negligibly till the system yields at a larger strain amplitude, beyond which the pressures are the same as for the poorly annealed glass. b) Pressure is linear with energy but increases with a larger slope beyond the yielding transition.} 
    \label{fig:pressInConstV}
\end{figure}

We begin by reporting the stroboscopic pressure values obtained for cyclic shear under constant volume conditions. Under constant volume simulations, the propensity for dilation upon yielding is manifested as an increase in the pressure beyond the yield point, as shown in \autoref{fig:pressInConstV}(a). For a well-annealed sample ($e_{IS}=-7.05$), pressure does not change for shear amplitudes less than the yield amplitude($\gamma^Y_{max}=0.10$). For a poorly annealed sample ($e_{IS}=-6.92$), the pressure decreases until the yield strain amplitude, beyond which the pressure follows a common curve with the well annealed sample. The observed behaviour closely follows the behaviour of the energy, which has been widely reported \cite{BhaumikPNAS2021}. The pressure is indeed found to be linearly correlated with energy, but with different slopes, before and after yielding,  as shown in \autoref{fig:pressInConstV}(b). For a large enough system \cite{fiocco2013}, yielding is associated with the formation of a shear band. Inside the shear band, the number density is found to be lower compared to the rest of the system \cite{ParmarPRX2019}. 

\begin{figure*}
    \centering
        \begin{tikzpicture}
            \node[anchor=south west, inner sep=0] (image) at (0,0) {\includegraphics[width=0.32\linewidth]{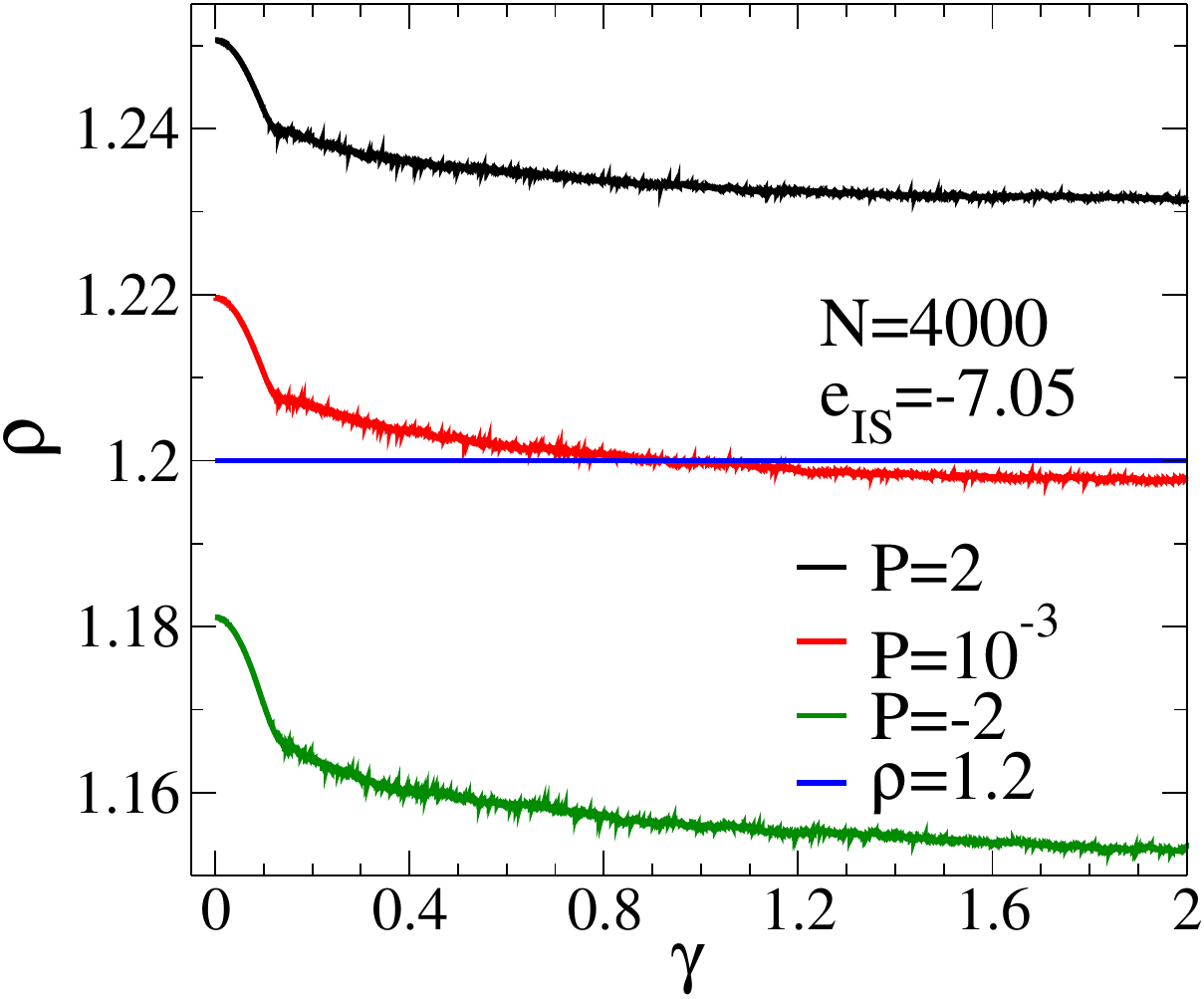}};
            \node[anchor=north west, font=\Large\bfseries] at (image.north west) [xshift=0mm, yshift=7mm] {a)};
        \end{tikzpicture}
    \hfill
        \begin{tikzpicture}
            \node[anchor=south west, inner sep=0] (image) at (0,0) {\includegraphics[width=0.32\linewidth]{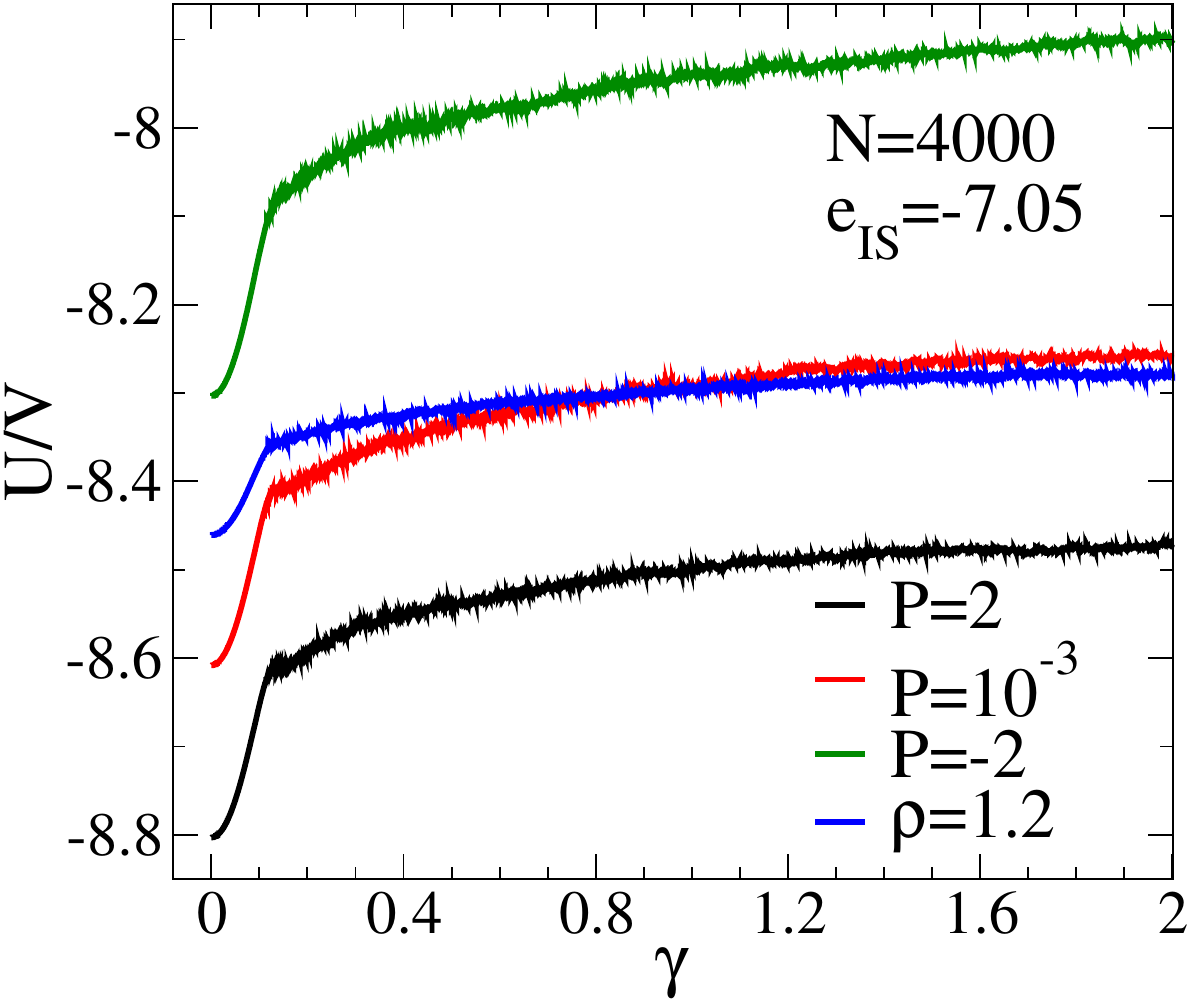}};
            \node[anchor=north west, font=\Large\bfseries] at (image.north west) [xshift=0mm, yshift=7mm] {b)};
        \end{tikzpicture}
        \hfill
        \begin{tikzpicture}
            \node[anchor=south west, inner sep=0] (image) at (0,0) {\includegraphics[width=0.32\linewidth]{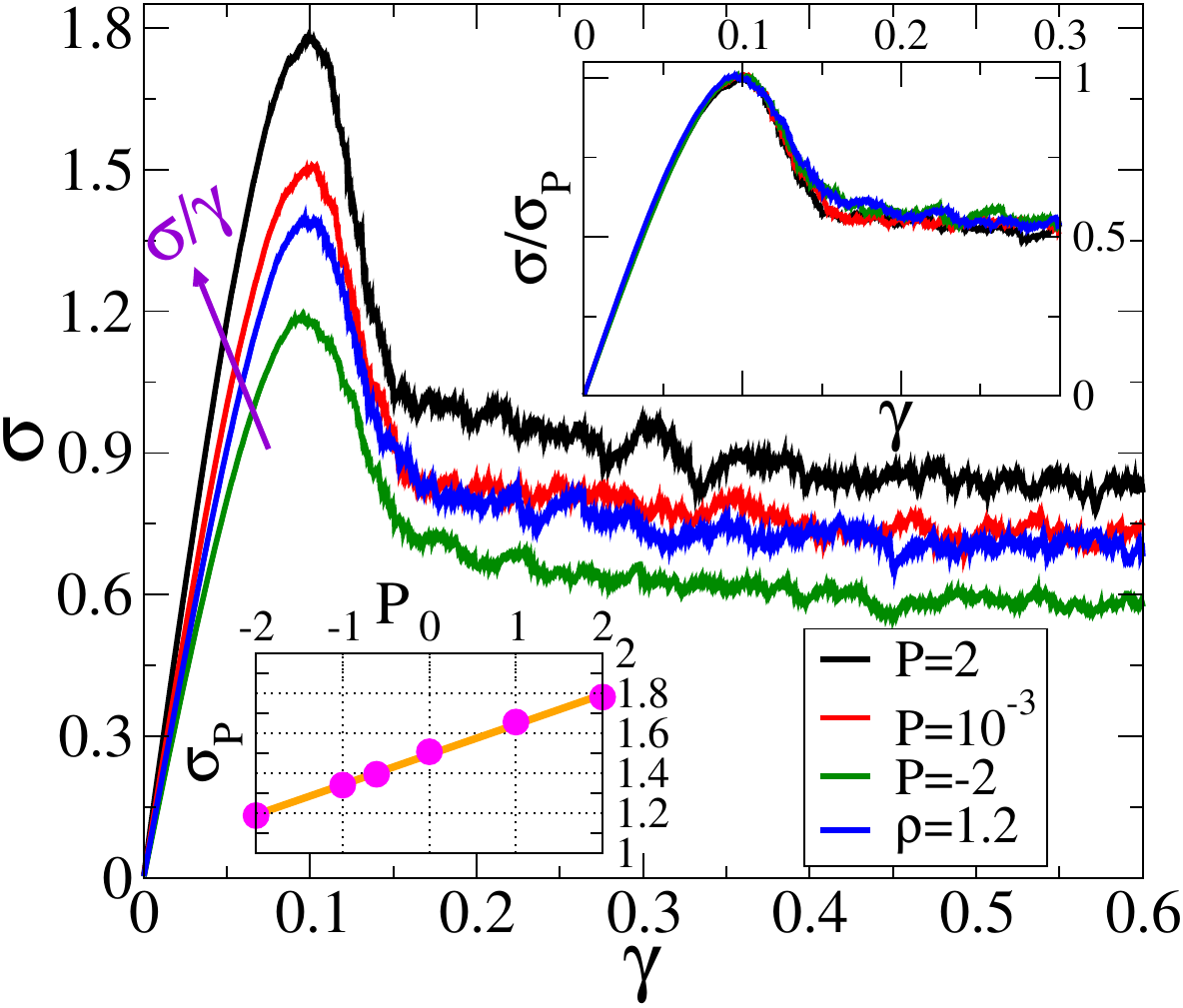}};
            \node[anchor=north west, font=\Large\bfseries] at (image.north west) [xshift=0mm, yshift=7mm] {c)};
        \end{tikzpicture}
        \\
        \begin{tikzpicture}
            \node[anchor=south west, inner sep=0] (image) at (0,0) {\includegraphics[width=0.315\linewidth]{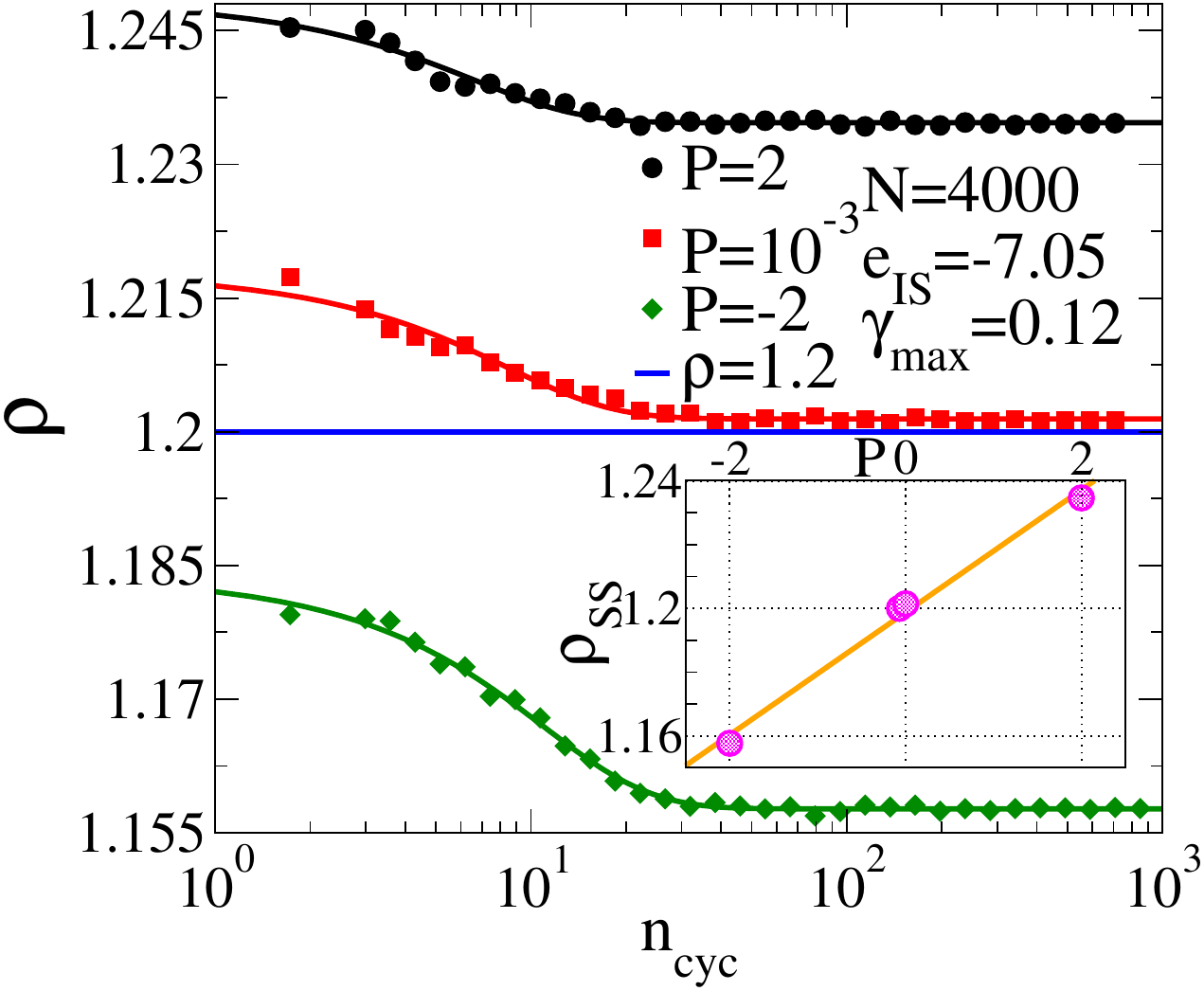}};
            \node[anchor=north west, font=\Large\bfseries] at (image.north west) [xshift=0mm, yshift=7mm] {d)};
        \end{tikzpicture}
    \hfill
        \begin{tikzpicture}
            \node[anchor=south west, inner sep=0] (image) at (0,0) {\includegraphics[width=0.32\linewidth]{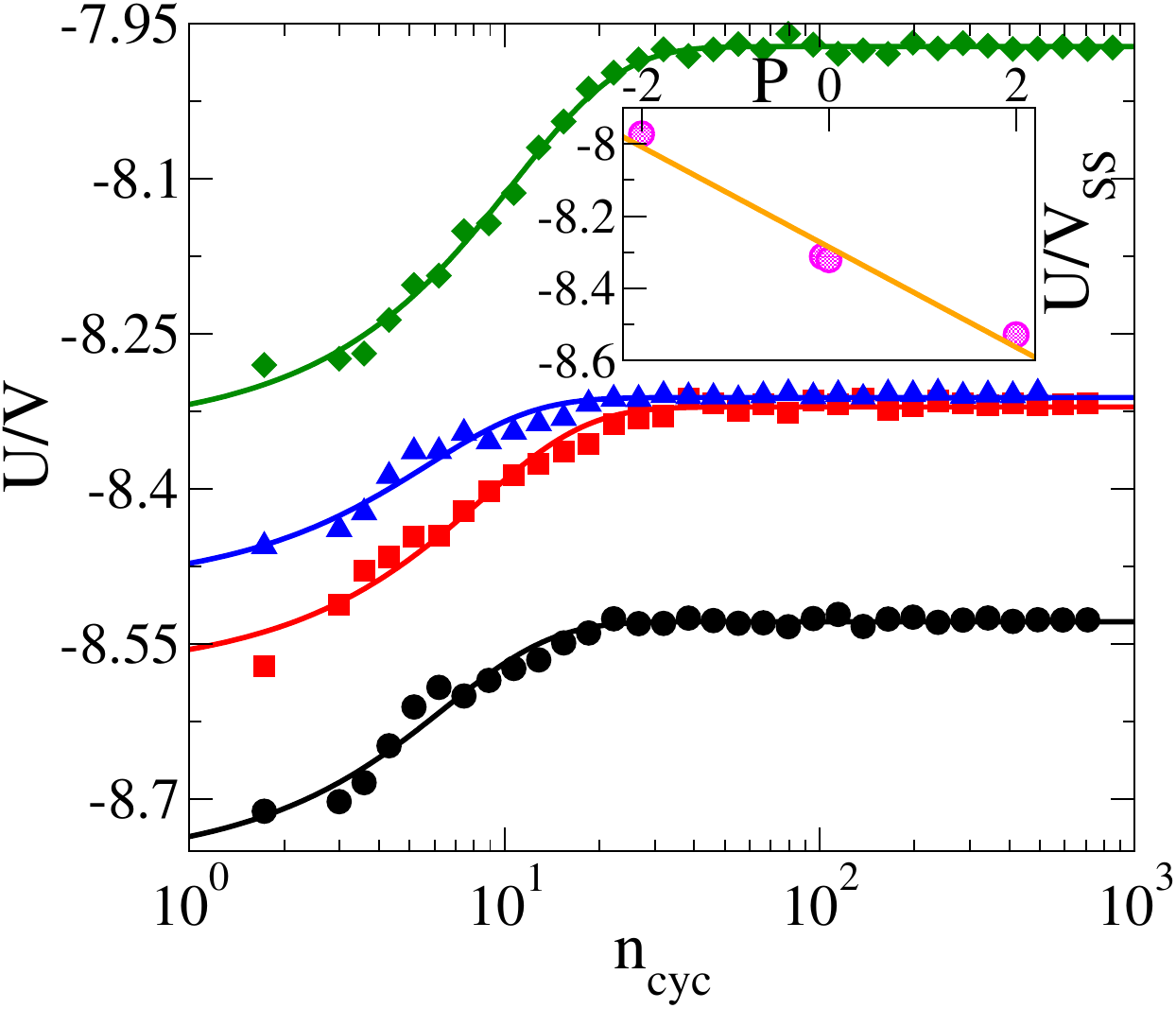}};
            \node[anchor=north west, font=\Large\bfseries] at (image.north west) [xshift=0mm, yshift=7mm] {e)};
        \end{tikzpicture}
        \hfill
        \begin{tikzpicture}
            \node[anchor=south west, inner sep=0] (image) at (0,0) {\includegraphics[width=0.32\linewidth]{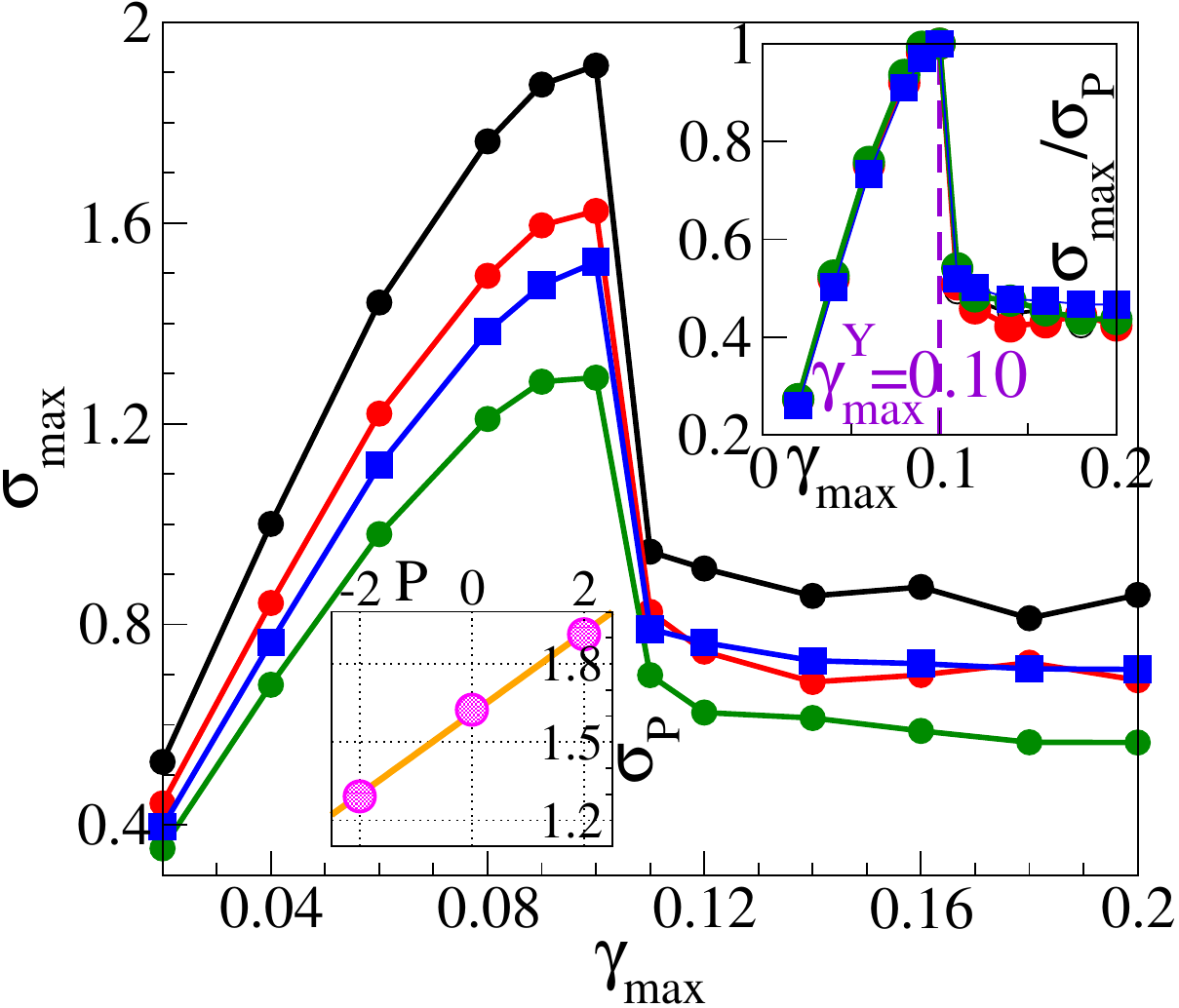}};
            \node[anchor=north west, font=\Large\bfseries] at (image.north west) [xshift=0mm, yshift=7mm] {f)};
        \end{tikzpicture}
    \caption{\label{fig:responses}\textbf{Density, Energy and Stress across the yielding transition} ($N = 4000$, $e_{IS} = -7.05$): \textbf{Uniform shear-} a) Evolution of number density with shear strain for a well-annealed glass at different external pressures under uniform shear. The density decreases both below and above the yield strain ($\gamma^Y \approx 0.1$).
    b) Energy density increases with strain, with the strain dependence changing at $\gamma^Y$.
 c) Stress $\sigma$ as a function of strain.  The peak stress ($\sigma_P$) before failure increases linearly with an increase in external pressure (shown in bottom left inset). All the stress-strain curves collapse on top of each other when stress values are rescaled with its peak value ($\sigma / \sigma_P$, top right inset). \textbf{Cyclic shear-} d) Evolution of number density with shear cycles at stroboscopic (at $\gamma=0$ after each cycle) configurations for strain amplitude $\gamma_{max} = 1.2$ for which failure happens in the first cycle. The system undergoes dilation and eventually reaches a steady state. The steady state number density ($\rho_{ss}$) increases linearly with increase in pressure (shown in the inset). The solid lines are fits to the data points used to obtain steady state values. e) The energy density $U/V$ of the system increases with cycles, and decreases with an increase in external pressure. The steady state value is linearly dependent on external pressure (as shown in the inset). f) The steady state value of stress at $\gamma = \gamma_{max}$ denoted by $\sigma_{max}$, increases with shear amplitude up to the yield point($\gamma_{max}^Y=0.10$). A sharp drop is observed for shear amplitudes greater than $\gamma_{max}^Y$ at all values of external pressures considered.  The solid lines are guides to the eye. The peak value of $\sigma_{max}$ increases linearly with external pressure (bottom left inset), and the $\sigma_{max}/\sigma_P$ are independent of constant volume or constant pressure cyclic shear conditions (top right inset).  }
\end{figure*}
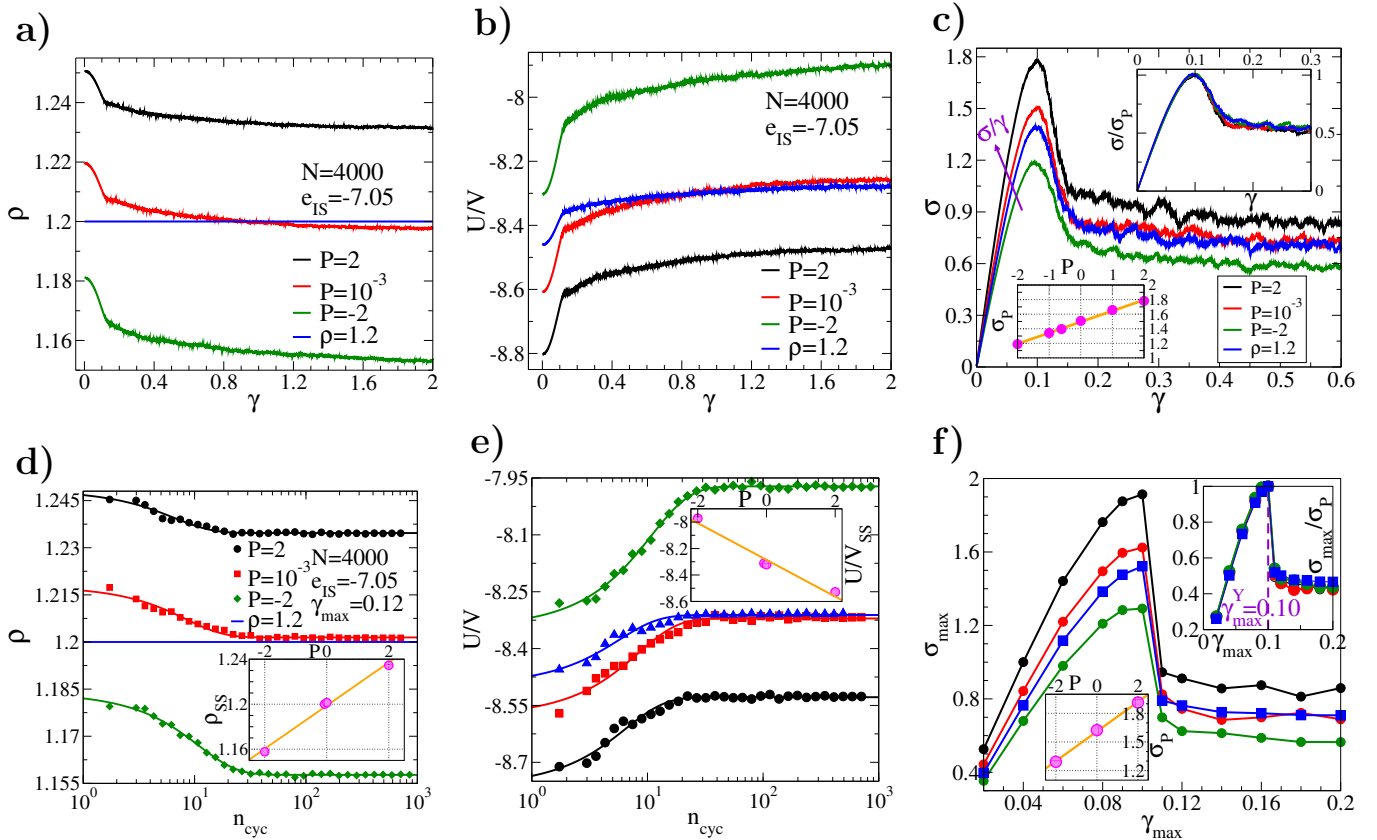

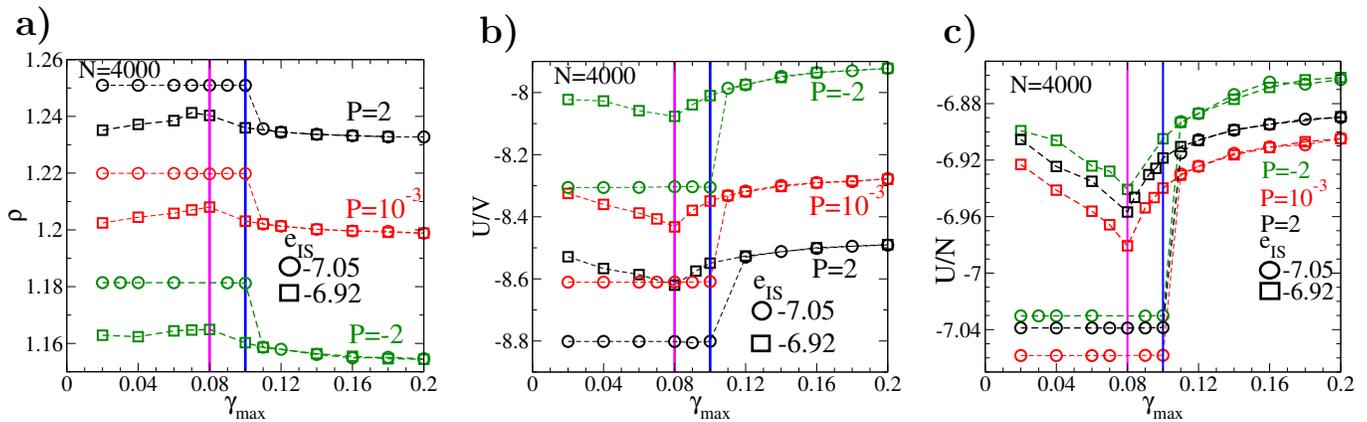
\begin{figure*}
    \centering
        \begin{tikzpicture}
            \node[anchor=south west, inner sep=0] (image) at (0,0) {\includegraphics[width=0.32\linewidth]{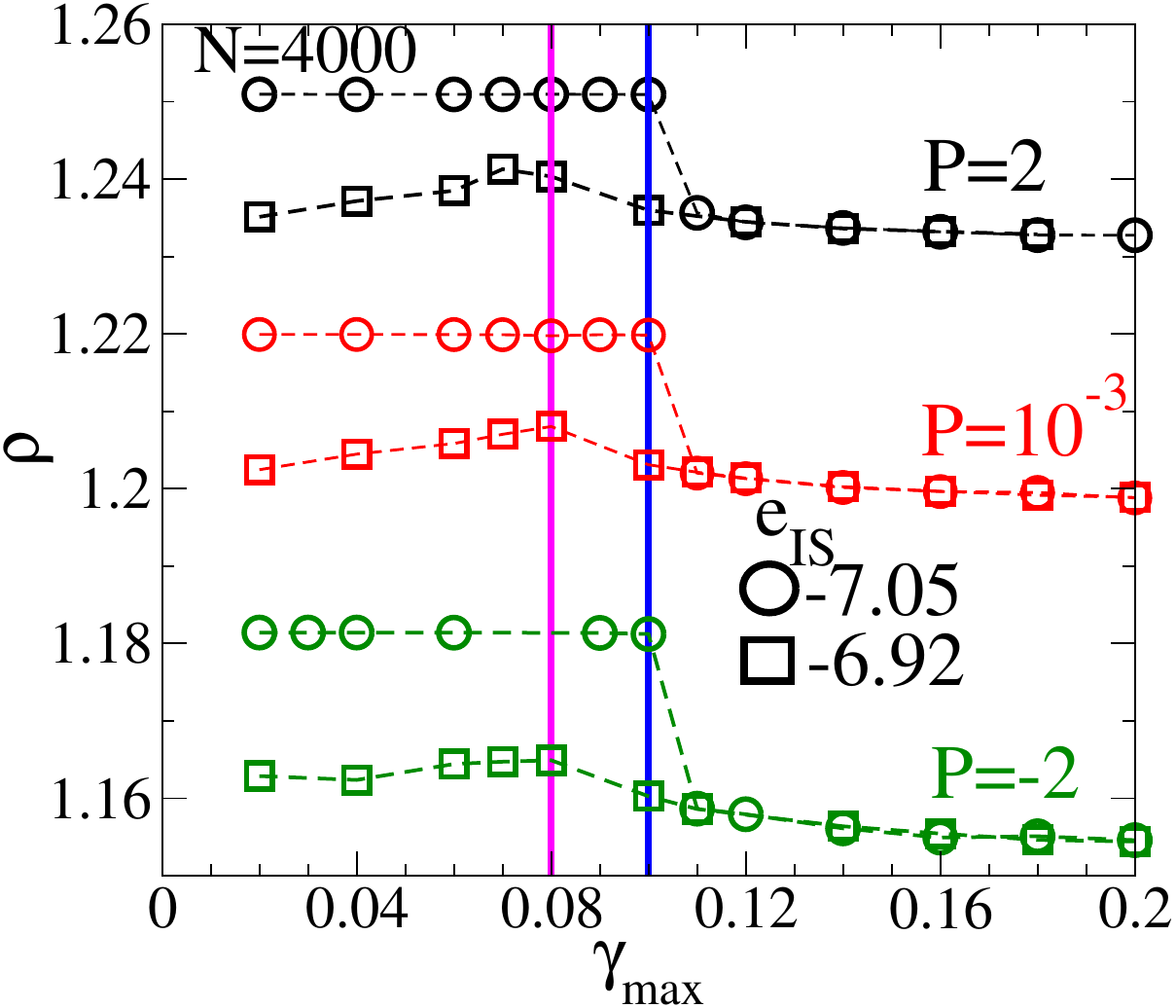}};
            \node[anchor=north west, font=\Large\bfseries] at (image.north west) [xshift=0mm, yshift=7mm] {a)};
        \end{tikzpicture}
    \hfill
        \begin{tikzpicture}
            \node[anchor=south west, inner sep=0] (image) at (0,0) {\includegraphics[width=0.32\linewidth]{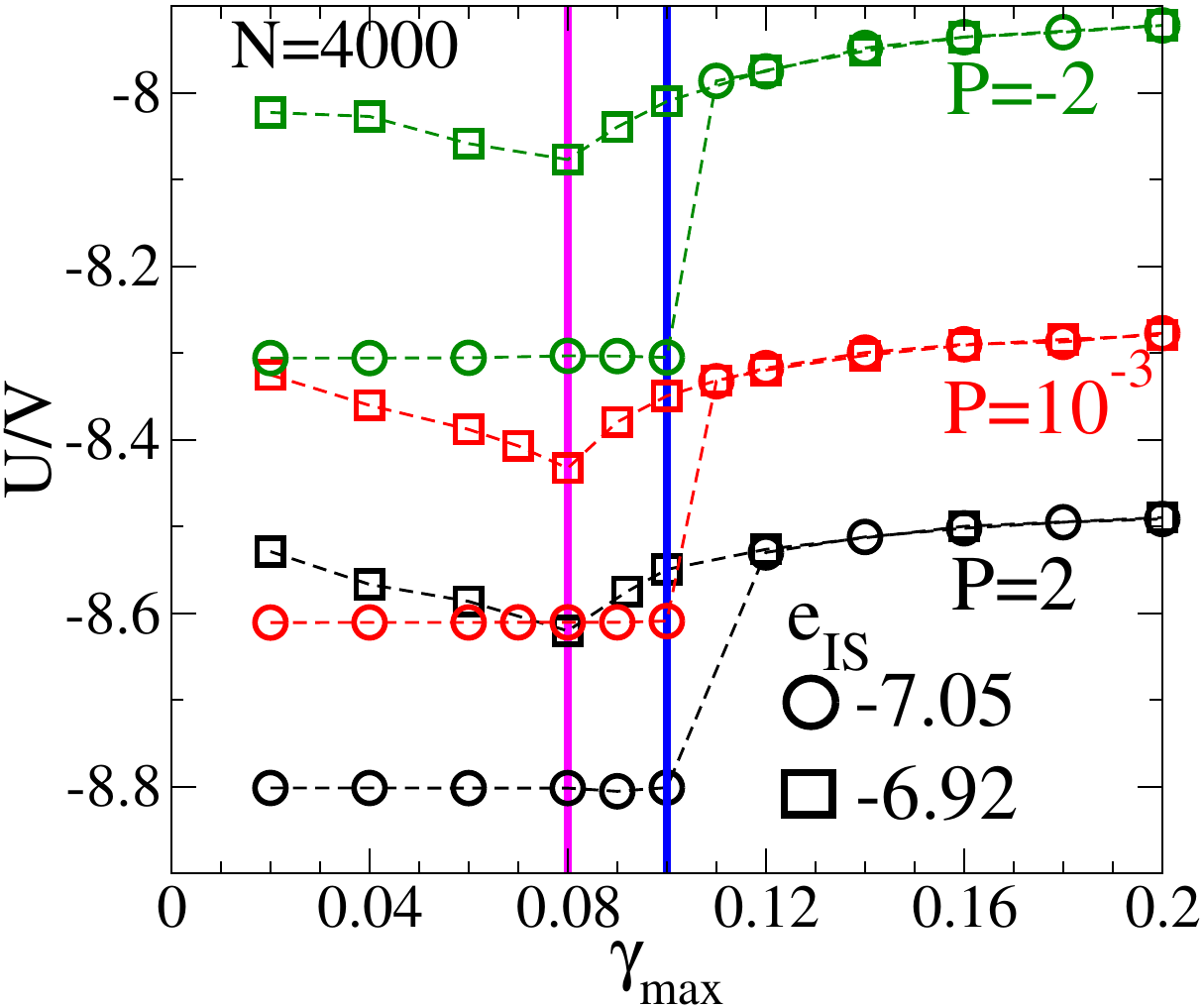}};
            \node[anchor=north west, font=\Large\bfseries] at (image.north west) [xshift=0mm, yshift=7mm] {b)};
        \end{tikzpicture}
    \hfill 
        \begin{tikzpicture}
            \node[anchor=south west, inner sep=0] (image) at (0,0) {\includegraphics[width=0.312\linewidth]{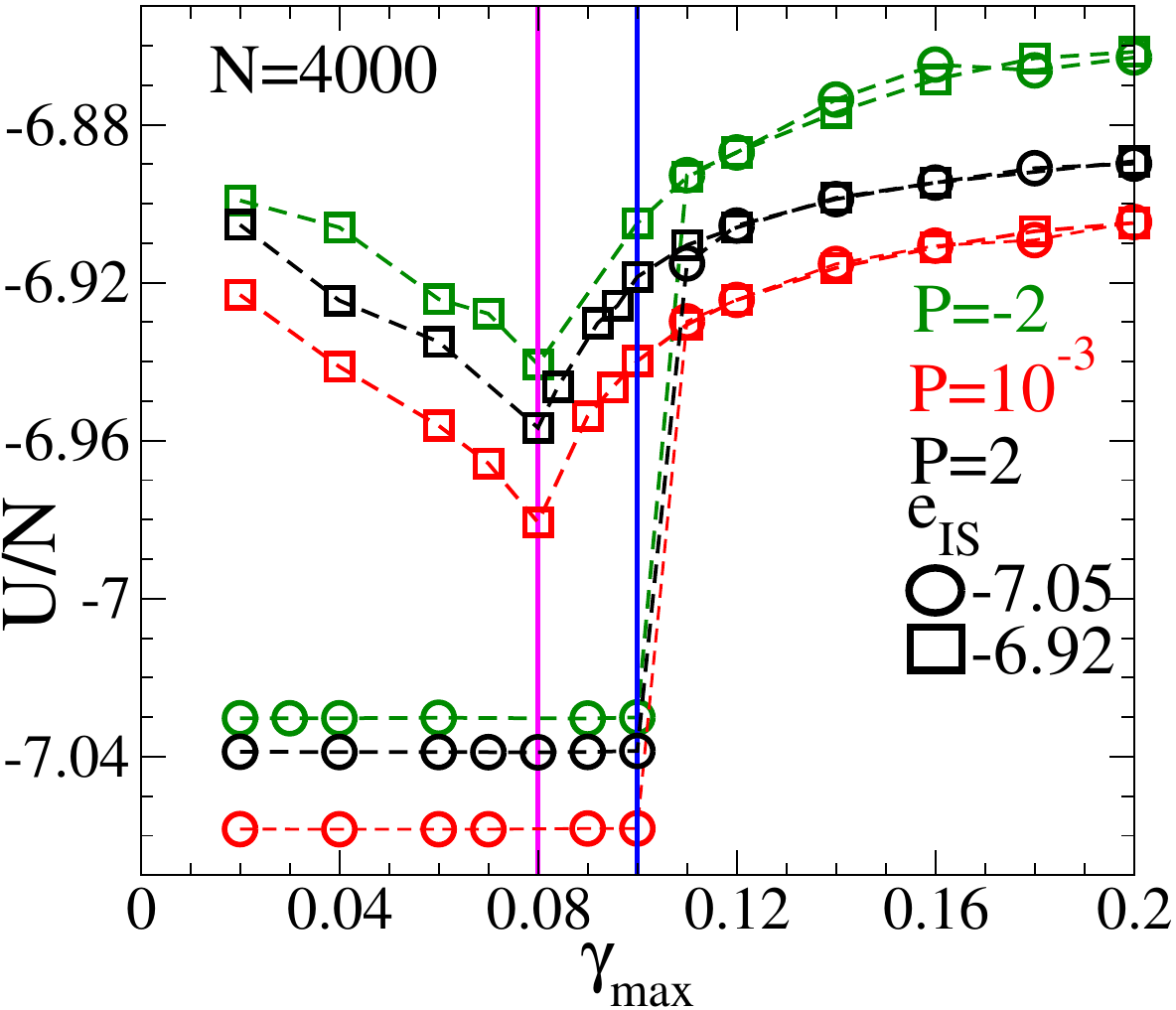}};
            \node[anchor=north west, font=\Large\bfseries] at (image.north west) [xshift=0mm, yshift=7mm] {c)};
        \end{tikzpicture}
    \caption{\textbf{Yielding diagrams at constant pressure:} The steady state values of number density $\rho$ and energy density $U/V$ are shown for poorly annealed ($e_{IS}=-6.92$, marked by open squares) and well annealed ($e_{IS}=-7.05$, marked by open circles) glasses. a) Number density: The poorly annealed glass undergoes compaction for small shear amplitudes followed by dilation after the yield point for all the values of pressure considered. The results for P=2 are shown in black, for P=$10^{-3}$ in red, and for P=$-2$ in green colors. The magenta and blue solid lines are the yield strain amplitudes for glasses with $e_{IS}=-6.92$ and $e_{IS}=-7.05$, respectively.  b) Energy density ($U/V$) {\it vs.} shear amplitude at different external pressures. The monotonic change in density dominates in the product of $\rho \times U/N = U/V$, resulting in a monotonic variation of $U/V$ with pressure. The energy density reflects the changes with annealing and strain amplitude seen in the number density and energy per particle.    
    c) Energy per particle ($U/N$): The poorly annealed glass undergoes annealing before the yield strain amplitude, followed by a discontinuous jump and subsequent increase with $\gamma_{max}$. The well annealed glass shows negligible annealing before yielding, and a larger discontinuous change at yielding. $U/N$ shows non-monotonic variation with pressure.} 
    \label{fig:yieldDiag}
\end{figure*}

\begin{figure*}
    \centering
        \begin{tikzpicture}
            \node[anchor=south west, inner sep=0] (image) at (0,0) {\includegraphics[width=0.32\linewidth]{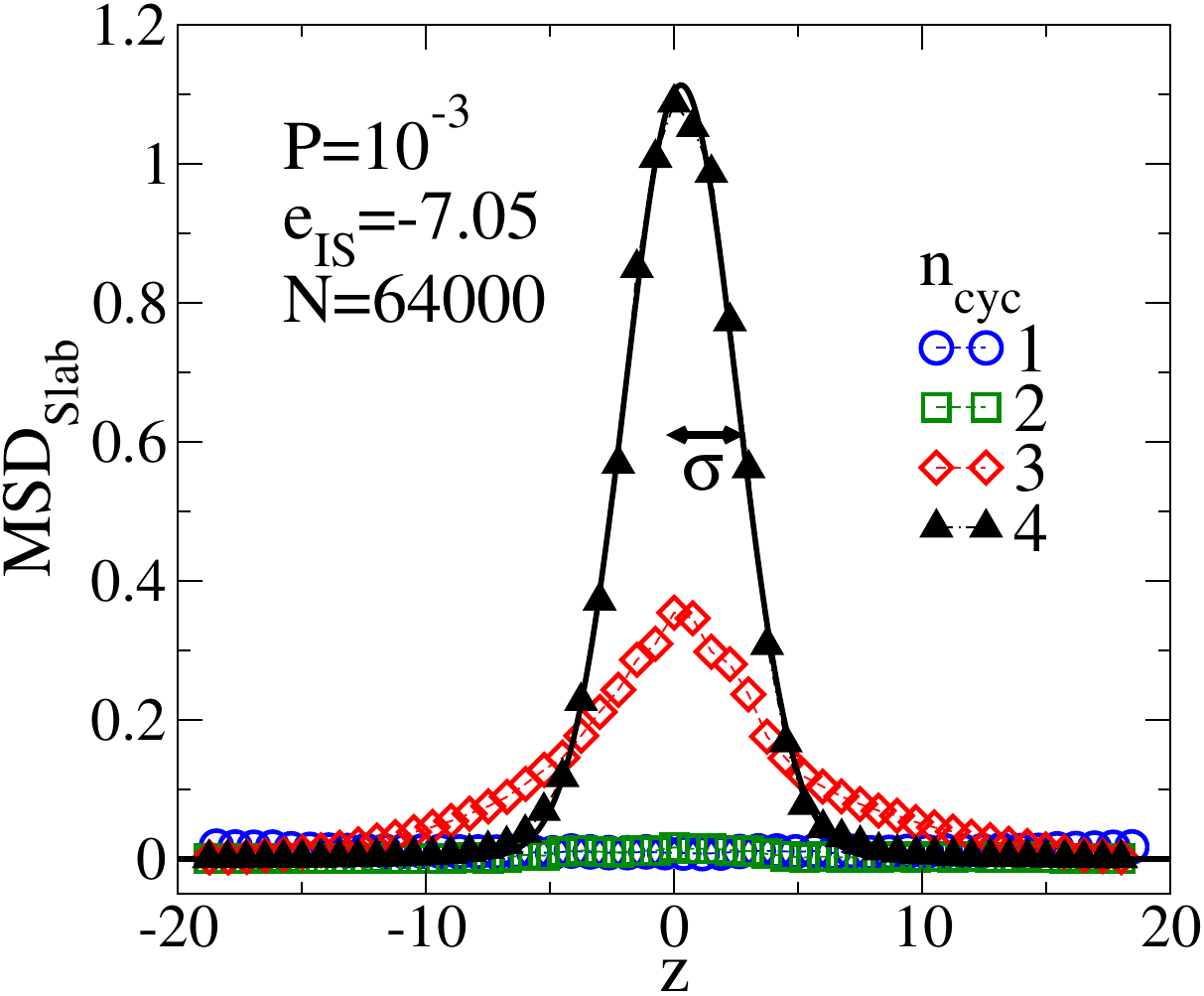}};
            \node[anchor=north west, font=\Large\bfseries] at (image.north west) [xshift=0mm, yshift=7mm] {a)};
        \end{tikzpicture}
    \hfill
        \begin{tikzpicture}
            \node[anchor=south west, inner sep=0] (image) at (0,0) {\includegraphics[width=0.32\linewidth]{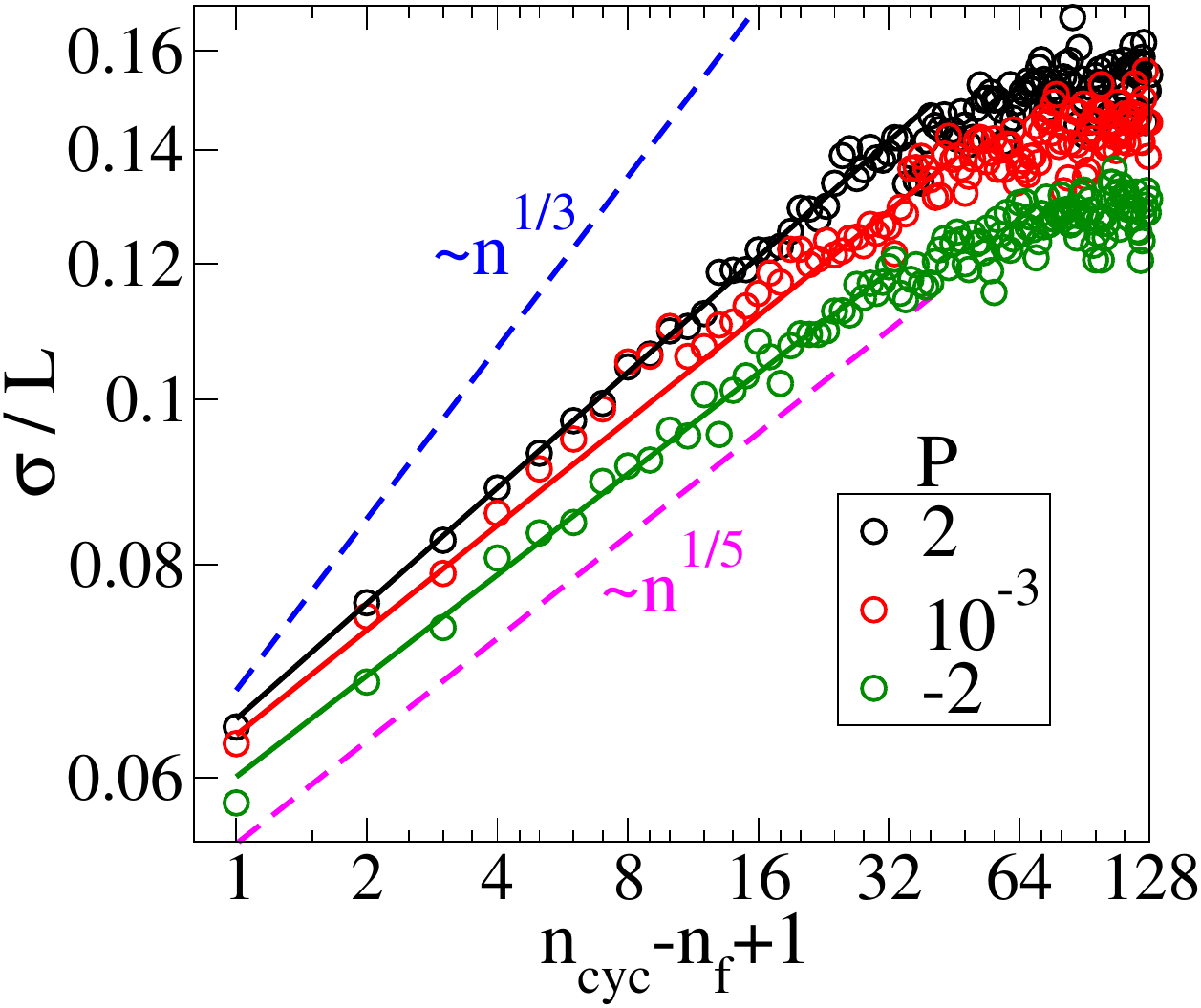}};
            \node[anchor=north west, font=\Large\bfseries] at (image.north west) [xshift=0mm, yshift=7mm] {b)};
        \end{tikzpicture}
        \hfill
        \begin{tikzpicture}
            \node[anchor=south west, inner sep=0] (image) at (0,0) {\includegraphics[width=0.32\linewidth]{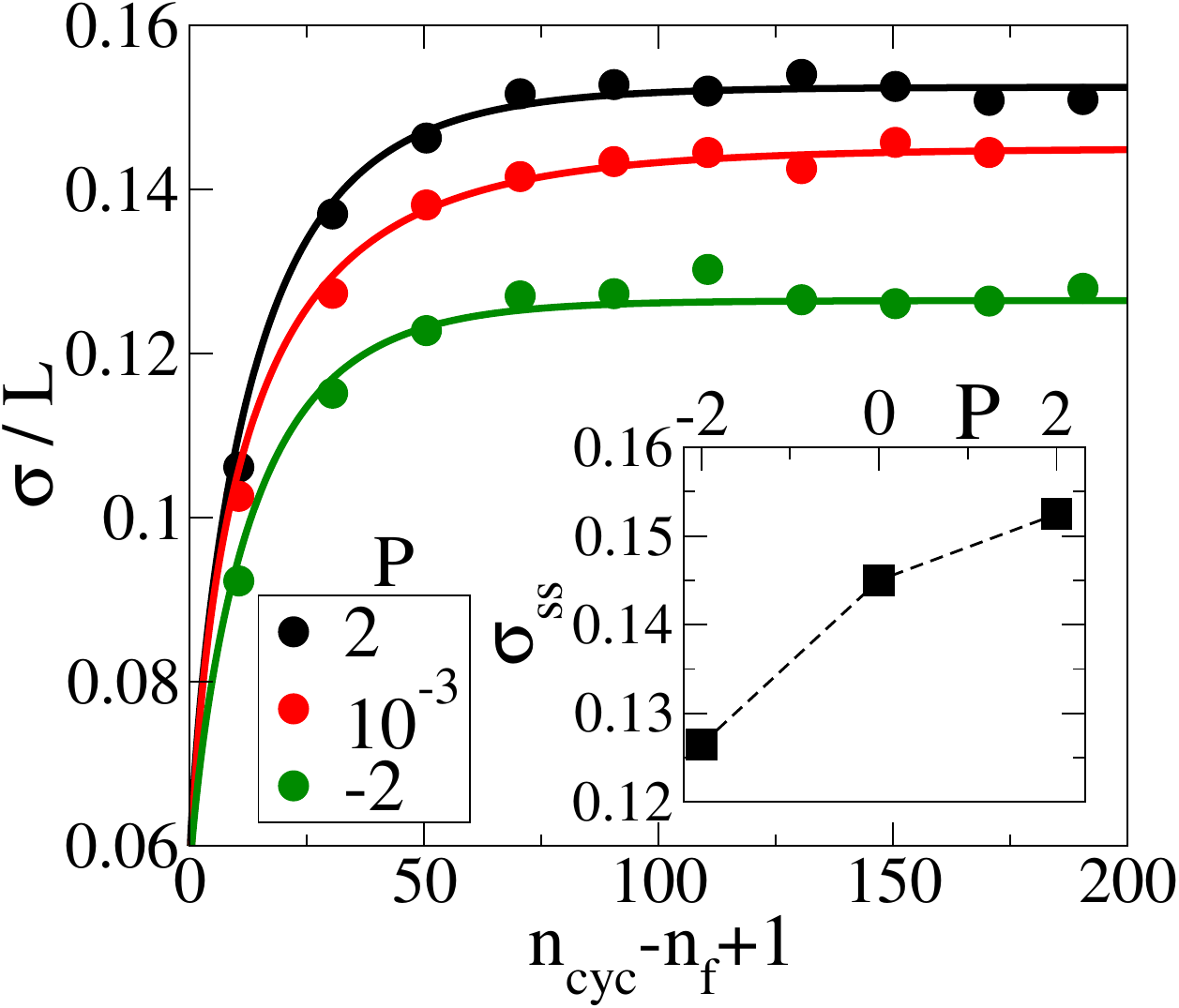}};
            \node[anchor=north west, font=\Large\bfseries] at (image.north west) [xshift=0mm, yshift=7mm] {c)};
        \end{tikzpicture}
        \\
        \begin{tikzpicture}
            \node[anchor=south west, inner sep=0] (image) at (0,0) {\includegraphics[width=0.32\linewidth]{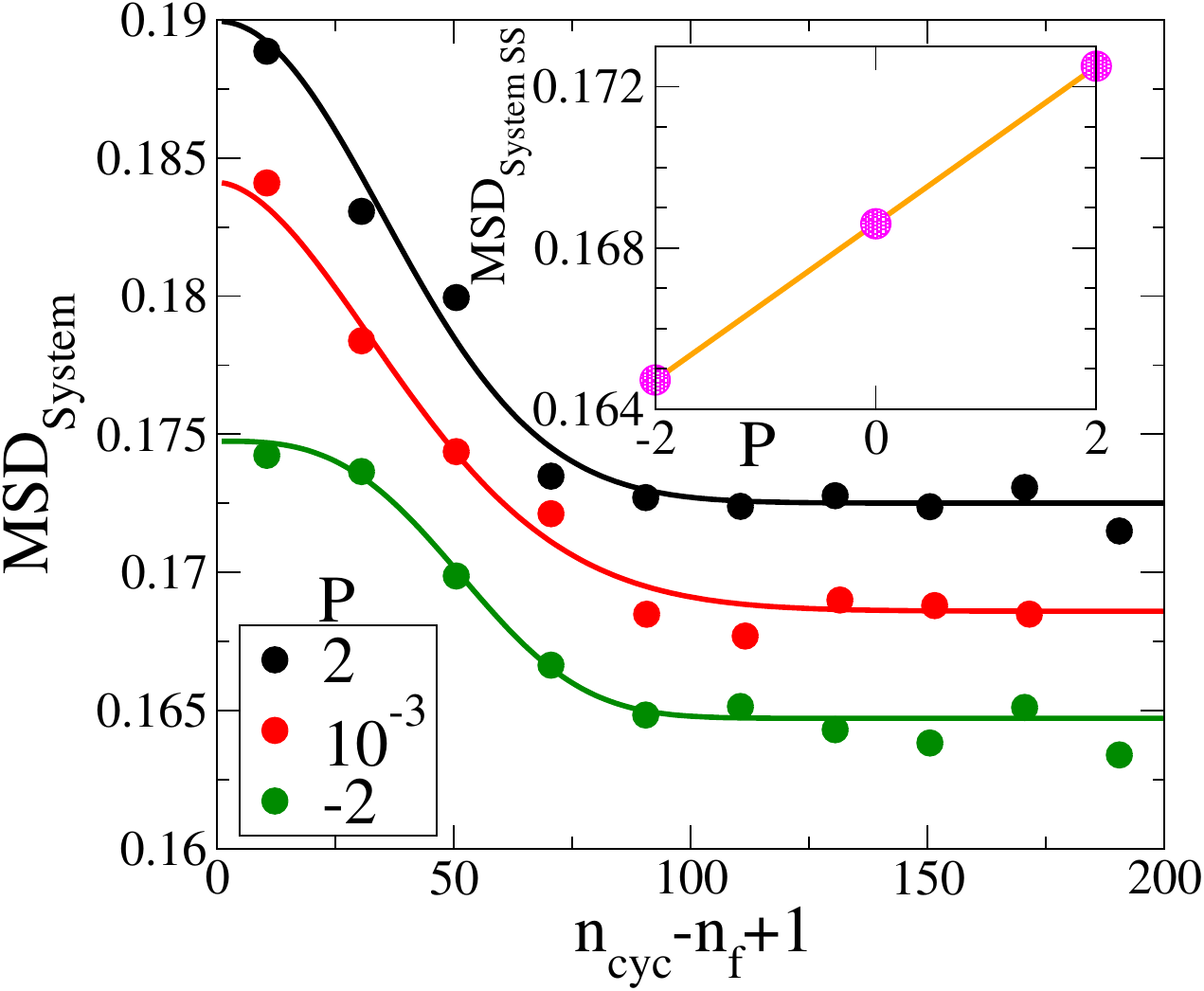}};
            \node[anchor=north west, font=\Large\bfseries] at (image.north west) [xshift=0mm, yshift=7mm] {d)};
        \end{tikzpicture}
    \hfill
        \begin{tikzpicture}
            \node[anchor=south west, inner sep=0] (image) at (0,0) {\includegraphics[width=0.32\linewidth]{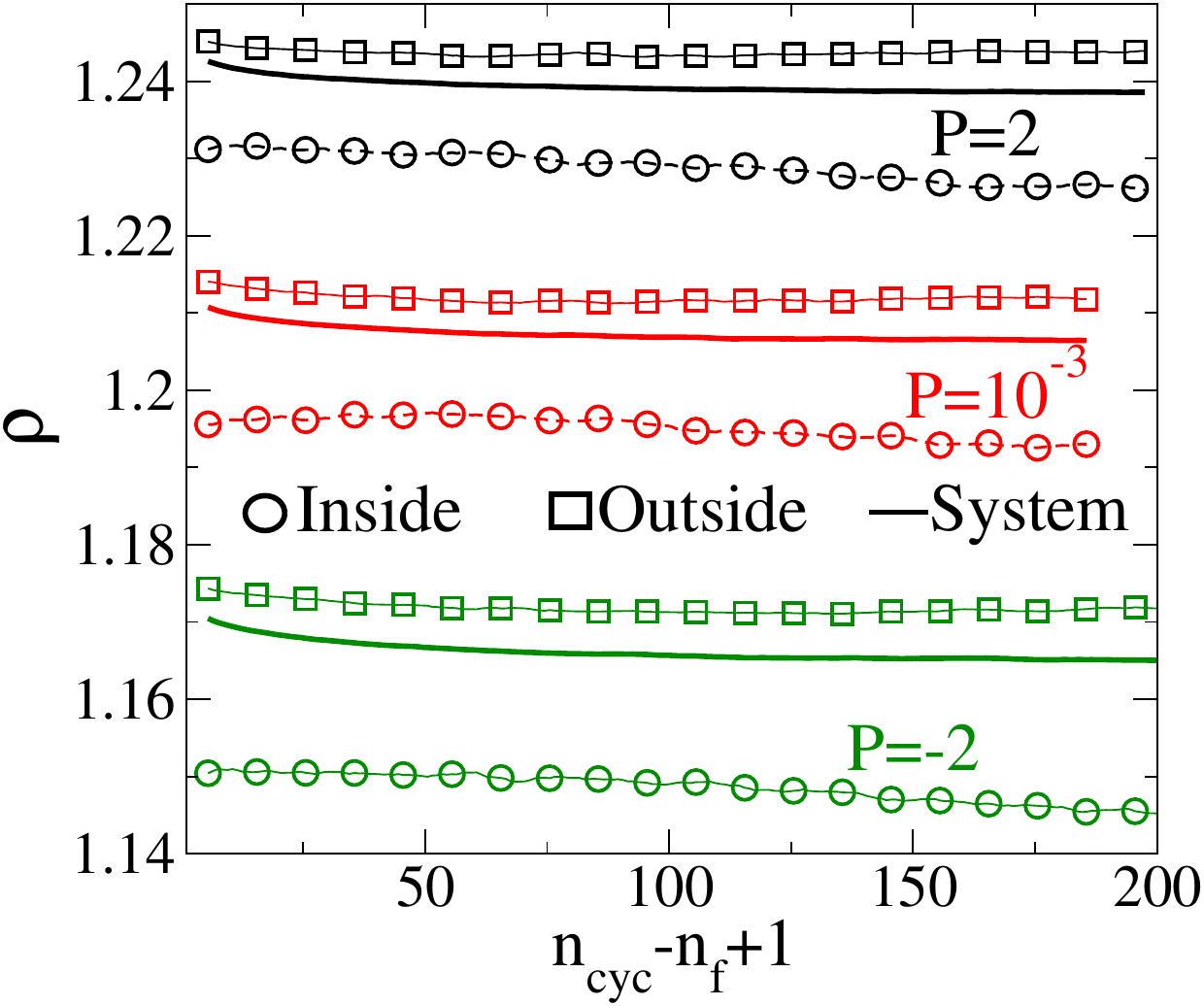}};
            \node[anchor=north west, font=\Large\bfseries] at (image.north west) [xshift=0mm, yshift=7mm] {e)};
        \end{tikzpicture}
        \hfill
        \begin{tikzpicture}
            \node[anchor=south west, inner sep=0] (image) at (0,0) {\includegraphics[width=0.32\linewidth]{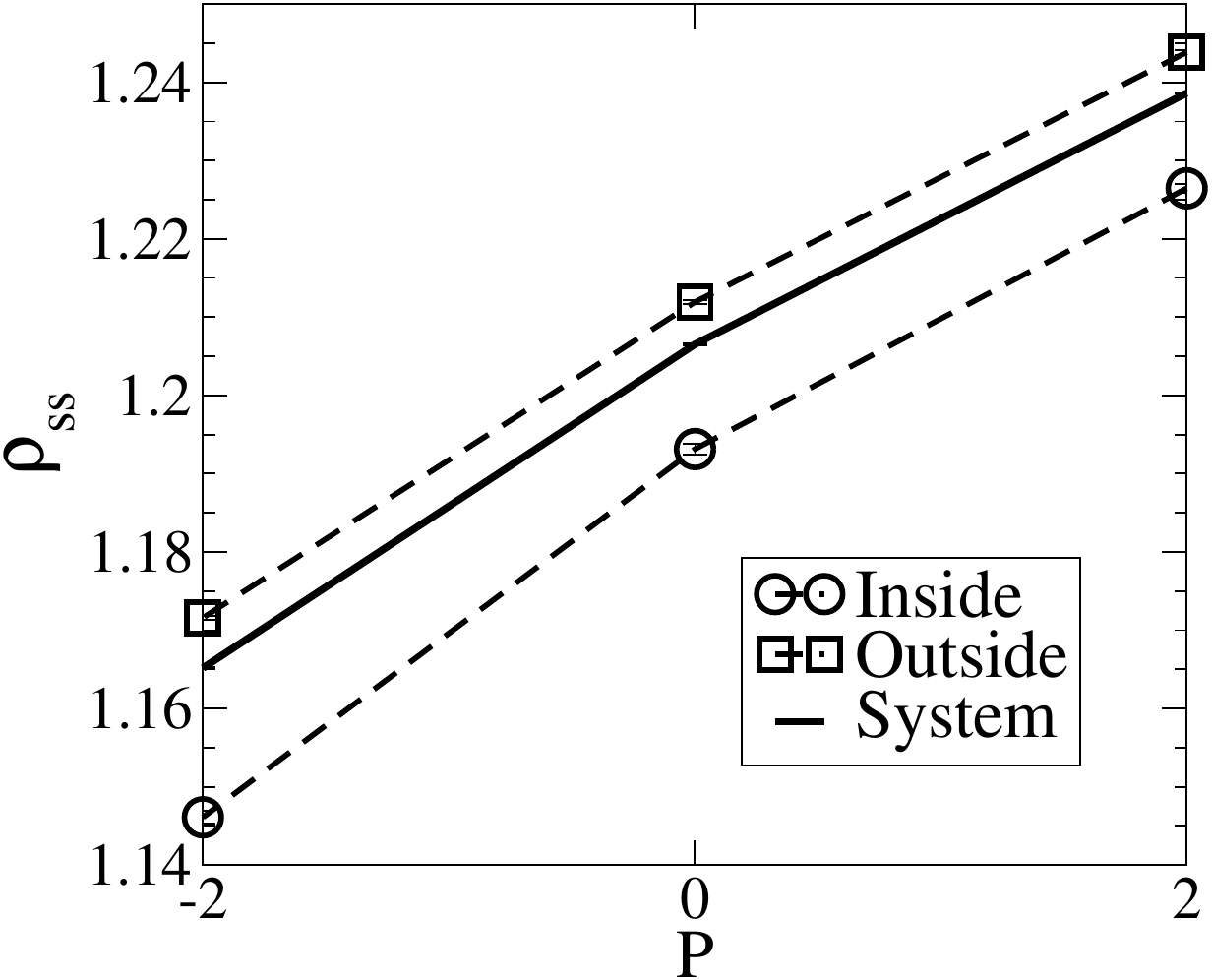}};
            \node[anchor=north west, font=\Large\bfseries] at (image.north west) [xshift=0mm, yshift=7mm] {f)};
        \end{tikzpicture}
    \caption{\label{fig:shearband}\textbf{Stability of shear bands at 
 constant pressure} (N=64000, $e_{IS}=-7.05$ and $\gamma_{max}=0.10$): a) Slab-wise cycle-to-cycle mean squared displacement profile ($MSD_{Slab}$) plotted as a function of the coordinate perpendicular to the shear band plane. We consider the glass to have yielded when the peak of $MSD_{Slab}$ is greater than $0.8$ (as a convenient but arbitrary choice), and denote the corresponding cycle by $n_f$. The $n_f$ value for the data shown is $4$. The shear band width $\sigma$ is obtained from a Gaussian fit to $MSD_{Slab}$, and dividing by the box length $L$ gives the fraction of the box length occupied by the shear band($\sigma / L$). b) Growth of the shear band with shear cycles after the formation of the shear band. The exponent of the power law growth is $\approx 0.2$, increasing mildly with pressure, and not close to $1/3$ as proposed in \cite{LiuJCP2022}.
  For reference, power laws with exponent $1/5$ and $1/3$ are shown with magenta and blue dashed lines respectively.  c) The shear band width increases and approaches a steady state value, being larger for larger pressure, as shown in the inset. The solid lines are fits to the binned data points used to obtain the steady state value. d) Mean squared displacement between subsequent stroboscopic configurations after the formation of the shear band ($MSD_{System}$) showing a high value at the formation of shear band, ultimately reaching a slightly lower steady state value. The steady state $MSD_{System}$ is higher for higher pressure. e) Variation of the number density with shear cycles inside (shown by open circle) and outside(shown by open squares) the shear band, and the system average value(shown by solid lines) at $P = 2$, $10^{-3}$ and $-2$ shown in black, red and green colors. The number density is lower inside the shear band compared to the average and the values outside. f) Steady state value of number density inside and outside of shear band as well as of the total number density of the system increases with pressure. } 
\end{figure*}



We next describe the results from constant pressure uniform and cyclic shear deformations. As stated earlier, we consider initial configurations (poorly annealed and well annealed), at three different pressures -- $P = 2, 10^{-3}, -2$. The variation of the density, energy density and stress are shown in \autoref{fig:responses}(a)-(c) for uniform shear and \autoref{fig:responses}(d)-(f) for cyclic shear, respectively. As shown later (in \autoref{fig:yieldDiag}), the energy per particle ($U/N$) shows non-monotonic dependence on pressure, whereas the energy density $U/V$ is monotonic, going to lower values as pressure increases. We show results for the well annealed case $e_{IS} = -7.05$ for samples with $N = 4000$. For uniform shear, the data are shown with respect to the applied strain. For cyclic shear, stroboscopic (zero strain) values of the density and energy density are shown as a function the number of cycles in \autoref{fig:responses}(d)-(e), whereas the maximum stress $\sigma_{max}$ reached through the cycle in the steady state is plotted as a function of the strain amplitude $\gamma_{max}$ in \autoref{fig:responses}(f). The results corresponding to constant volume shear at $\rho = 1.2$ are also shown for comparison. 

\autoref{fig:responses}(a) shows that the number density under uniform shear decreases in all cases, both before and after yielding, with the densities being higher for larger pressures. The energy density in  \autoref{fig:responses}(b) reflects the same trends, however with the energy density being the lowest for the highest pressure, and with all values increasing with strain. The stress-strain curves in \autoref{fig:responses}(c) show a clear signatures of yielding, with a stress overshoot as expected for a well annealed glass. The slope of the stress-strain curve, as well as the peak stress $\sigma_P$, increases with pressure. Interestingly, scaling the stress values with the peak stress $\sigma_P$ leads to a collapse of all the cases shown, including the case of constant volume shear.  Under cyclic shear, as shown in \autoref{fig:responses}(d) for a case for which the system fails in the first cycle, the (stroboscopic) number density $\rho$ values decrease with the cycles of shear, and reach steady state values that increase with increasing pressure. The energy density shows the opposite behaviour, increasing with cycles, and the steady state values show a decrease with increasing pressure. Finally, the maximum stress $\sigma_{max}$ increases with strain amplitude $\gamma_{max}$ till the yield value, at which it displays a discontinuous drop in all cases. As in the case of uniform shear, scaling the $\sigma_{max}$ values with the peak value leads to a data collapse of results for all the cases including the constant volume cyclic shear.

The steady state values of the number density, energy density and energy per particle of the system sheared at different shear amplitudes under constant pressure are shown for different pressures in \autoref{fig:yieldDiag}. The number densities increased compaction with increase in strain amplitude $\gamma_{max}$ before yielding, whereas well annealed glasses show negligible change. Beyond the yield point, both poorly and well annealed glasses show a reduction in the density with increased strain amplitude $\gamma_{max}$. The variation of the energy density $U/V$ and energy $U/N$, shown in \autoref{fig:yieldDiag} (b) and (c), qualitatively follow behaviour in  constant volume yielding diagrams\cite{BhaumikPNAS2021}. The modifications due to pressure are in shifting the energy curves. The energy density shifts to lower values with an increase in pressure as shown in \autoref{fig:yieldDiag} (b). The yield point does not depend on the external pressure.


Yielding of amorphous systems is accompanied by the formation of shear bands within which  strain gets localized. The transition is also accompanied by the displacement of particles from one cycle to the next, leading to diffusive motion over repeated cycles. The cycle to cycle displacements (referred to as $MSD$ below), when averaged over slabs parallel to the shear banding plane, are sharply peaked within the shear band, and are reasonably well fitted by a Gaussian. Following \cite{ParmarPRX2019}, we fit a Gaussian function to the slab-averaged $MSD$ and use the standard deviation of the fit as a measure of the width of the shear band,  as shown in \autoref{fig:shearband}(a) for a large system of $N = 64000$ particles, at $\gamma_{max} = 0.1$. The width of the shear band as a function of the number of cycles after formation, exhibits a power law growth, as shown in \autoref{fig:shearband}(b). It has been suggested \cite{LiuJCP2022} that the coarsening dynamics of shear band growth is well described by an exponent of $1/3$ ({\it i. e.}, the width $w \sim n^{1/3}$ where $n$ is the number of cycles). While we find a power law growth, we do not find an exponent value of $1/3$ for any pressure, as shown in  \autoref{fig:shearband}(b). The shear band width eventually saturates, as shown in \autoref{fig:shearband}(c), with steady state values that increase with the pressure. A primary conclusion, therefore, is that the shear bands remain stable despite the dilatancy observed, when the global volume of the glass is allowed to vary. 

As the shear band forms, $MSD$  values in the shear-banded region increase, followed by a decrease and a widening of the shear band. The resulting system averaged $MSD$  values are shown in \autoref{fig:shearband}(d), which indicate that the cycle-to-cycle displacements of particles also reach stable values, at larger values for larger pressures, consistently with the larger width. The previous observation of lowered number densities within the shear band \cite{ParmarPRX2019} persists also in the constant pressure simulations we perform, as shown in \autoref{fig:shearband}(e) and (f). The densities inside and outside the shear bands increase with pressure, as does the average density. 

\section{Summary and Discussion}
In summary, we have investigated the yielding behaviour of an amorphous solid subjected to strain-controlled shear simulations at constant external pressure. Specifically, we investigated the stability of the shear bands accompanying yielding under cyclic shear deformation, in view of previous observations of a reduction in density within the shear band \cite{ParmarPRX2019} under constant volume conditions, and find that shear bands indeed remain stable under constant pressure conditions. We find that qualitatively, the phenomenology of yielding under constant pressure conditions largely tracks the behaviour seen 
in constant volume simulations. For both uniform and cyclic shear deformation, a change in pressure scales the stress values observed, with a scaling of the stress values leading to data collapse across the different pressure, as well as constant volume, shear protocols. Thus, overall, we conclude that the phenomenology of uniform and cyclic shear yielding is not qualitatively affected by the relaxation of the fixed volume condition by applying strain at constant pressure conditions. Nevertheless, there are several ways in which further investigations along these lines are warranted. The first is the occurrence of dilatational effects under shear strain conditions itself,  a surprising feature that is, beyond broad qualitative arguments, not well investigated and understood. Most theoretical modeling efforts to understand yielding focus on the shear strain protocol and ignore volumetric effects. Second, the volumetric effects appear to also  involve compositional segregation (unpublished results), which may be of interest to understand better. Finally, the important special case of silica exhibits compaction, rather than dilatation, upon shear banding \cite{BhaumikPRL2022}. This feature appears associated with the anomalous relationship between energetics and volume in the case of silica. Therefore, it is of interest to understand dilatancy and failure in a broader framework that includes such cases. 

\textit{Acknowledgements: } We thank H. Bhaumik, A. L. Greer, A. Lemaitre and A. Pandey for useful discussions and comments on the manuscript. S. S. acknowledges SERB(ANRF) (India) for support through the JC Bose Fellowship (JBR/2020/000015) SERB(ANRF), DST (India) and a grant under SUPRA (SPR/2021/000382). K.K.T acknowledge the financial support from the Council of Scientific and Industrial Research (CSIR), India, through the Junior/Senior Research Fellowship (Award No. 09/733(0285)/2020-EMR I).




\bibliography{main.bib}

@article{BhaumikPRL2022,
  title = {Avalanches, Clusters, and Structural Change in Cyclically Sheared Silica Glass},
  author = {Bhaumik, Himangsu and Foffi, Giuseppe and Sastry, Srikanth},
  journal = {Phys. Rev. Lett.},
  volume = {128},
  issue = {9},
  pages = {098001},
  numpages = {6},
  year = {2022},
  month = {Feb},
  publisher = {American Physical Society},
  doi = {10.1103/PhysRevLett.128.098001},
  url = {https://link.aps.org/doi/10.1103/PhysRevLett.128.098001}
}

@inbook{Parmar2019b,
author = {Anshul D. S. Parmar and Srikanth Sastry},
title = {Mechanical Behaviour of Glasses and Amorphous Materials},
booktitle = {Advances in the Chemistry and Physics of Materials},
chapter = {Chapter 21},
pages = {503-527},
doi = {10.1142/9789811211331_0021},
URL = {https://www.worldscientific.com/doi/abs/10.1142/9789811211331_0021}
}

@article{Reynolds1885,
author = {Osborne Reynolds},
title = {LVII. On the dilatancy of media composed of rigid particles in contact. With experimental illustrations },
journal = {The London, Edinburgh, and Dublin Philosophical Magazine and Journal of Science},
volume = {20},
number = {127},
pages = {469--481},
year = {1885},
publisher = {Taylor \& Francis},
doi = {10.1080/14786448508627791},


URL = { 
    
        https://doi.org/10.1080/14786448508627791
    
    

},
eprint = { 
    
        https://doi.org/10.1080/14786448508627791
}
}

@article{Ren2013,
  title = {Reynolds Pressure and Relaxation in a Sheared Granular System},
  author = {Ren, Jie and Dijksman, Joshua A. and Behringer, Robert P.},
  journal = {Phys. Rev. Lett.},
  volume = {110},
  issue = {1},
  pages = {018302},
  numpages = {5},
  year = {2013},
  month = {Jan},
  publisher = {American Physical Society},
  doi = {10.1103/PhysRevLett.110.018302},
  url = {https://link.aps.org/doi/10.1103/PhysRevLett.110.018302}
}

@Article{Babu2021,
author ="Babu, Varghese and Pan, Deng and Jin, Yuliang and Chakraborty, Bulbul and Sastry, Srikanth",
title  ="Dilatancy{,} shear jamming{,} and a generalized jamming phase diagram of frictionless sphere packings",
journal  ="Soft Matter",
year  ="2021",
volume  ="17",
issue  ="11",
pages  ="3121-3127",
publisher  ="The Royal Society of Chemistry",
doi  ="10.1039/D0SM02186E",
url  ="http://dx.doi.org/10.1039/D0SM02186E"}

@article{Jiang2017,
  author = {Jiang, M. Q. and Wilde, G. and Dai, L. H.},
  title = {Shear band dilatation in amorphous alloys},
  journal = {Scripta Materialia},
  year = {2017},
  volume = {127},
  pages = {54--57},
  doi = {10.1016/j.scriptamat.2016.08.038},
  url = {https://doi.org/10.1016/j.scriptamat.2016.08.038}
}

@article{Sun2019,
  author = {Sun, X. and Ding, G. and Mo, G. and Dai, L. H. and Jiang, M. Q.},
  title = {Dilatancy signatures of amorphous plasticity probed by X-ray synchrotron radiation},
  journal = {Intermetallics},
  year = {2019},
  volume = {107},
  pages = {34--38},
  doi = {10.1016/j.intermet.2019.01.009},
  url = {https://doi.org/10.1016/j.intermet.2019.01.009}
}

@article{Wang2016,
  author = {Wang, Yun-Jiang and Jiang, M. Q. and Tian, Z. L. and Dai, L. H.},
  title = {Direct atomic-scale evidence for shear-dilatation correlation in metallic glasses},
  journal = {Scripta Materialia},
  year = {2016},
  volume = {112},
  pages = {37--41},
  doi = {10.1016/j.scriptamat.2015.09.005},
  url = {https://doi.org/10.1016/j.scriptamat.2015.09.005}
}

@article{Wang2021,
  author = {Wang, Xiaodi and Wu, Shaojie and Qu, Ruitao and Zhang, Zhefeng},
  title = {Shear Band Evolution under Cyclic Loading and Fatigue Property in Metallic Glasses: A Brief Review},
  journal = {Materials},
  year = {2021},
  volume = {14},
  number = {13},
  pages = {3595},
  doi = {10.3390/ma14133595},
  url = {https://doi.org/10.3390/ma14133595}
}

@article{Lu2018,
  author = {Lu, Y. Z. and Jiang, M. Q. and Lu, X. and Qin, Z. X. and Huang, Y. J. and Shen, J.},
  title = {Dilatancy of Shear Transformations in a Colloidal Glass},
  journal = {Physical Review Applied},
  year = {2018},
  volume = {9},
  number = {1},
  pages = {014023},
  doi = {10.1103/PhysRevApplied.9.014023},
  url = {https://doi.org/10.1103/PhysRevApplied.9.014023}
}

@article{Lemaitre2002,
  author = {Lemaître, Anaël},
  title = {Rearrangements and Dilatancy for Sheared Dense Materials},
  journal = {Physical Review Letters},
  year = {2002},
  volume = {89},
  number = {19},
  pages = {195503},
  doi = {10.1103/PhysRevLett.89.195503},
  url = {https://doi.org/10.1103/PhysRevLett.89.195503}
}

@article{Moriel2024,
  author = {Moriel, Avraham and Richard, David and Lerner, Edan and Bouchbinder, Eran},
  title = {Elementary processes in dilatational plasticity of glasses},
  journal = {Physical Review Research},
  year = {2024},
  volume = {6},
  number = {2},
  pages = {023167},
  doi = {10.1103/PhysRevResearch.6.023167},
  url = {https://doi.org/10.1103/PhysRevResearch.6.023167}
}

@article{Zeng2018,
  author = {Zeng, F. and Jiang, M. Q. and Dai, L. H.},
  title = {Dilatancy induced ductile-brittle transition of shear band in metallic glasses},
  journal = {Proceedings of the Royal Society A: Mathematical, Physical and Engineering Sciences},
  year = {2018},
  volume = {474},
  number = {2213},
  pages = {20170836},
  doi = {10.1098/rspa.2017.0836},
  url = {https://doi.org/10.1098/rspa.2017.0836}
}

@article{BonnRevModPhys2017,
  title = {Yield stress materials in soft condensed matter},
  author = {Bonn, Daniel and Denn, Morton M. and Berthier, Ludovic and Divoux, Thibaut and Manneville, S\'ebastien},
  journal = {Rev. Mod. Phys.},
  volume = {89},
  issue = {3},
  pages = {035005},
  numpages = {40},
  year = {2017},
  month = {Aug},
  publisher = {American Physical Society},
  doi = {10.1103/RevModPhys.89.035005},
  url = {https://link.aps.org/doi/10.1103/RevModPhys.89.035005}
}

@article{NicolasRevModPhys2018,
  title = {Deformation and flow of amorphous solids: Insights from elastoplastic models},
  author = {Nicolas, Alexandre and Ferrero, Ezequiel E. and Martens, Kirsten and Barrat, Jean-Louis},
  journal = {Rev. Mod. Phys.},
  volume = {90},
  issue = {4},
  pages = {045006},
  numpages = {63},
  year = {2018},
  month = {Dec},
  publisher = {American Physical Society},
  doi = {10.1103/RevModPhys.90.045006},
  url = {https://link.aps.org/doi/10.1103/RevModPhys.90.045006}
}

@article{parmar2019,
  title = {Strain Localization Above the Yielding Point in Cyclically Deformed Glasses},
  author = {Parmar, Anshul D. S. and Kumar, Saurabh and Sastry, Srikanth},
  journal = {Phys. Rev. X},
  volume = {9},
  issue = {2},
  pages = {021018},
  numpages = {11},
  year = {2019},
  month = {Apr},
  publisher = {American Physical Society},
  doi = {10.1103/PhysRevX.9.021018},
  url = {https://link.aps.org/doi/10.1103/PhysRevX.9.021018}
}

@article{das_parmar_22,
    author = {Das, Pallabi and Parmar, Anshul D. S. and Sastry, Srikanth},
    title = {Annealing glasses by cyclic shear deformation},
    journal = {The Journal of Chemical Physics},
    volume = {157},
    number = {4},
    pages = {044501},
    year = {2022},
    month = {08},
    issn = {0021-9606},
    doi = {10.1063/5.0100523},
    url = {https://doi.org/10.1063/5.0100523}
}

@article{SunPRL10,
  title = {Plasticity of Ductile Metallic Glasses: A Self-Organized Critical State},
  author = {Sun, B. A. and Yu, H. B. and Jiao, W. and Bai, H. Y. and Zhao, D. Q. and Wang, W. H.},
  journal = {Phys. Rev. Lett.},
  volume = {105},
  issue = {3},
  pages = {035501},
  numpages = {4},
  year = {2010},
  month = {Jul},
  publisher = {American Physical Society},
  doi = {10.1103/PhysRevLett.105.035501},
  url = {https://link.aps.org/doi/10.1103/PhysRevLett.105.035501}
}

@article{AntonagliaPRL14,
  title = {Bulk Metallic Glasses Deform via Slip Avalanches},
  author = {Antonaglia, James and Wright, Wendelin J. and Gu, Xiaojun and Byer, Rachel R. and Hufnagel, Todd C. and LeBlanc, Michael and Uhl, Jonathan T. and Dahmen, Karin A.},
  journal = {Phys. Rev. Lett.},
  volume = {112},
  issue = {15},
  pages = {155501},
  numpages = {5},
  year = {2014},
  month = {Apr},
  publisher = {American Physical Society},
  doi = {10.1103/PhysRevLett.112.155501},
  url = {https://link.aps.org/doi/10.1103/PhysRevLett.112.155501}
}

@Article{maloneypre06,
  title		= {Amorphous systems in athermal, quasistatic shear},
  author	= {Maloney, Craig E. and Lema\^{\i}tre, Ana\"el},
  journal	= {Phys. Rev. E},
  volume	= {74},
  issue		= {1},
  pages		= {016118},
  numpages	= {22},
  year		= {2006},
  month		= {Jul},
  publisher	= {American Physical Society},
  doi		= {10.1103/PhysRevE.74.016118},
  url		= {https://link.aps.org/doi/10.1103/PhysRevE.74.016118}
}

@Article{karmakarpre10,
  title		= {Statistical physics of the yielding transition in amorphous solids},
  author	= {Karmakar, Smarajit and Lerner, Edan and Procaccia, Itamar},
  journal	= {Phys. Rev. E},
  volume	= {82},
  issue		= {5},
  pages		= {055103},
  numpages	= {4},
  year		= {2010},
  month		= {Nov},
  publisher	= {American Physical Society},
  doi		= {10.1103/PhysRevE.82.055103},
  url		= {https://link.aps.org/doi/10.1103/PhysRevE.82.055103}
}

@Article{jaiswalprl16,
  title		= {Mechanical Yield in Amorphous Solids: A First-Order Phase Transition},
  author	= {Jaiswal, Prabhat K. and Procaccia, Itamar and Rainone, Corrado and Singh, Murari},
  journal	= {Phys. Rev. Lett.},
  volume	= {116},
  issue		= {8},
  pages		= {085501},
  numpages	= {5},
  year		= {2016},
  month		= {Feb},
  publisher	= {American Physical Society},
  doi		= {10.1103/PhysRevLett.116.085501},
  url		= {https://link.aps.org/doi/10.1103/PhysRevLett.116.085501}
}

@Article{fiocco2013,
  title		= {Oscillatory athermal quasistatic deformation of a model glass},
  author	= {Fiocco, Davide and Foffi, Giuseppe and Sastry, Srikanth},
  journal	= {Phys. Rev. E},
  volume	= {88},
  issue		= {2},
  pages		= {020301},
  numpages	= {5},
  year		= {2013},
  month		= {Aug},
  publisher	= {American Physical Society},
  doi		= {10.1103/PhysRevE.88.020301},
  url		= {https://link.aps.org/doi/10.1103/PhysRevE.88.020301}
}

@Article{priezjev2013,
  title		= {Heterogeneous relaxation dynamics in amorphous materials under cyclic loading},
  author	= {Priezjev, Nikolai V.},
  journal	= {Phys. Rev. E},
  volume	= {87},
  issue		= {5},
  pages		= {052302},
  numpages	= {6},
  year		= {2013},
  month		= {May},
  publisher	= {American Physical Society},
  doi		= {10.1103/PhysRevE.87.052302},
  url		= {https://link.aps.org/doi/10.1103/PhysRevE.87.052302}
}

@Article{regev2015reversibility,
  title		= {Reversibility and criticality in amorphous solids},
  author	= {Regev, Ido and Weber, John and Reichhardt, Charles and Dahmen, Karin A and Lookman, Turab},
  journal	= {Nature Communications},
  volume	= {6},
  number	= {1},
  pages		= {8805--8805},
  year		= {2015}
}

@Article{leishangthemNAT2017,
  author	= {Leishangthem, Premkumar and Parmar, Anshul D. S. and Sastry, Srikanth},
  title		= {The yielding transition in amorphous solids under oscillatory shear deformation},
  journal	= {Nature Communications},
  year		= {2017},
  month		= {Mar},
  day		= {01},
  publisher	= {The Author(s)},
  volume	= {8},
  pages		= {14653},
  url		= {http://dx.doi.org/10.1038/ncomms14653}
}

@Article{Jin,
  title		= {A stability-reversibility map unifies elasticity, plasticity, yielding and jamming in hard sphere glasses},
  author	= {Jin, Yuliang and Urbani, Pierfrancesco and Zamponi, Francesco and Yoshino, Hajime},
  journal	= {Science Advances},
  volume	= {4},
  number	= {12},
  year		= {2018}
}

@Article{ozawapnas2018,
  doi		= {10.1073/pnas.1806156115},
  url		= {https://doi.org/10.1073/pnas.1806156115},
  year		= {2018},
  month		= jun,
  publisher	= {Proceedings of the National Academy of Sciences},
  volume	= {115},
  number	= {26},
  pages		= {6656--6661},
  author	= {Misaki Ozawa and Ludovic Berthier and Giulio Biroli and Alberto Rosso and Gilles Tarjus},
  title		= {Random critical point separates brittle and ductile yielding transitions in amorphous materials},
  journal	= {Proceedings of the National Academy of Sciences}
}

@Article{barbotpre20,
  title		= {Rejuvenation and shear banding in model amorphous solids},
  author	= {Barbot, Armand and Lerbinger, Matthias and Lema\^{\i}tre, Ana\"el and Vandembroucq, Damien and Patinet, Sylvain},
  journal	= {Phys. Rev. E},
  volume	= {101},
  issue		= {3},
  pages		= {033001},
  numpages	= {10},
  year		= {2020},
  month		= {Mar},
  publisher	= {American Physical Society},
  doi		= {10.1103/PhysRevE.101.033001},
  url		= {https://link.aps.org/doi/10.1103/PhysRevE.101.033001}
}

@article{Dasgupta2012,
title = {Microscopic Mechanism of Shear Bands in Amorphous Solids},
  author = {Dasgupta, Ratul and Hentschel, H. George E. and Procaccia, Itamar},
  journal = {Phys. Rev. Lett.},
  volume = {109},
  issue = {25},
  pages = {255502},
  numpages = {4},
  year = {2012},
  month = {Dec},
  publisher = {American Physical Society},
  doi = {10.1103/PhysRevLett.109.255502},
  url = {https://link.aps.org/doi/10.1103/PhysRevLett.109.255502}
}

@Article{LinPNAS14,
  author	= {Lin, Jie and Lerner, Edan and Rosso, Alberto and Wyart,
		  Matthieu},
  title		= {Scaling description of the yielding transition in soft
		  amorphous solids at zero temperature},
  volume	= {111},
  number	= {40},
  pages		= {14382--14387},
  year		= {2014},
  doi		= {10.1073/pnas.1406391111},
  publisher	= {National Academy of Sciences},
  issn		= {0027-8424},
  url		= {https://www.pnas.org/content/111/40/14382},
  journal	= {Proceedings of the National Academy of Sciences}
}

@Article{parisipnas17,
  author	= {Parisi, Giorgio and Procaccia, Itamar and Rainone, Corrado
		  and Singh, Murari},
  title		= {Shear bands as manifestation of a criticality in yielding
		  amorphous solids},
  volume	= {114},
  number	= {22},
  pages		= {5577--5582},
  year		= {2017},
  doi		= {10.1073/pnas.1700075114},
  publisher	= {National Academy of Sciences},
  issn		= {0027-8424},
  url		= {https://www.pnas.org/content/114/22/5577},
  journal	= {Proceedings of the National Academy of Sciences}
}

@Article{urbani2017b,
  title		= {Shear Yielding and Shear Jamming of Dense Hard Sphere
		  Glasses},
  author	= {Urbani, Pierfrancesco and Zamponi, Francesco},
  journal	= {Physical Review Letters},
  volume	= {118},
  number	= {3},
  pages		= {38001},
  year		= {2017}
}

@Article{BudrikisNATCOM2017,
author={Budrikis, Zoe
and Castellanos, David Fernandez
and Sandfeld, Stefan
and Zaiser, Michael
and Zapperi, Stefano},
title={Universal features of amorphous plasticity},
journal={Nature Communications},
year={2017},
month={Jul},
day={03},
volume={8},
number={1},
pages={15928},
doi={10.1038/ncomms15928},
url={https://doi.org/10.1038/ncomms15928}
}

@Article{popovic2018a,
  title		= {Elastoplastic description of sudden failure in athermal
		  amorphous materials during quasistatic loading},
  author	= {Popovi{\'c}, Marko and de Geus, Tom WJ and Wyart,
		  Matthieu},
  journal	= {Physical Review E},
  volume	= {98},
  number	= {6},
  pages		= {040901},
  year		= {2018}
}

@Article{barlow2020,
  title		= {Ductile and Brittle Yielding in Thermal and Athermal
		  Amorphous Materials},
  author	= {Barlow, Hugh J. and Cochran, James O. and Fielding,
		  Suzanne M.},
  journal	= {Phys. Rev. Lett.},
  volume	= {125},
  issue		= {16},
  pages		= {168003},
  numpages	= {6},
  year		= {2020},
  month		= {Oct},
  publisher	= {American Physical Society},
  doi		= {10.1103/PhysRevLett.125.168003},
  url		= {https://link.aps.org/doi/10.1103/PhysRevLett.125.168003}
}

@article{liu2020oscillatory,
    author = {Liu, Chen and Ferrero, Ezequiel E. and Jagla, Eduardo A. and Martens, Kirsten and Rosso, Alberto and Talon, Laurent},
    title = {The fate of shear-oscillated amorphous solids},
    journal = {The Journal of Chemical Physics},
    volume = {156},
    number = {10},
    pages = {104902},
    year = {2022},
    month = {03},
    issn = {0021-9606},
    doi = {10.1063/5.0079460},
    url = {https://doi.org/10.1063/5.0079460}
}

@Article{sastryPRL20,
  title = {Models for the Yielding Behavior of Amorphous Solids},
  author = {Sastry, Srikanth},
  journal = {Phys. Rev. Lett.},
  volume = {126},
  issue = {25},
  pages = {255501},
  numpages = {5},
  year = {2021},
  month = {Jun},
  publisher = {American Physical Society},
  doi = {10.1103/PhysRevLett.126.255501},
  url = {https://link.aps.org/doi/10.1103/PhysRevLett.126.255501}
}

@Article{khirallahPRL20,
  title = {Yielding in an Integer Automaton Model for Amorphous Solids under Cyclic Shear},
  author = {Khirallah, Kareem and Tyukodi, Botond and Vandembroucq, Damien and Maloney, Craig E.},
  journal = {Phys. Rev. Lett.},
  volume = {126},
  issue = {21},
  pages = {218005},
  numpages = {6},
  year = {2021},
  month = {May},
  publisher = {American Physical Society},
  doi = {10.1103/PhysRevLett.126.218005},
  url = {https://link.aps.org/doi/10.1103/PhysRevLett.126.218005}
}

@article{ParleyPhysFluid2020,
    author = {Parley, Jack T. and Fielding, Suzanne M. and Sollich, Peter},
    title = {Aging in a mean field elastoplastic model of amorphous solids},
    journal = {Physics of Fluids},
    volume = {32},
    number = {12},
    pages = {127104},
    year = {2020},
    month = {12},
    issn = {1070-6631},
    doi = {10.1063/5.0033196},
    url = {https://doi.org/10.1063/5.0033196}
}

@Article{MunganPRL2021,
  title = {Metastability as a Mechanism for Yielding in Amorphous Solids under Cyclic Shear},
  author = {Mungan, Muhittin and Sastry, Srikanth},
  journal = {Phys. Rev. Lett.},
  volume = {127},
  issue = {24},
  pages = {248002},
  numpages = {6},
  year = {2021},
  month = {Dec},
  publisher = {American Physical Society},
  doi = {10.1103/PhysRevLett.127.248002},
  url = {https://link.aps.org/doi/10.1103/PhysRevLett.127.248002}
}

@misc{sarkar2025,
      title={Coarse grained descriptions of the dynamics of yielding of amorphous solids under cyclic shear}, 
      author={Debargha Sarkar and Jishnu N. Nampoothiri and Muhittin Mungan and Jack T. Parley and Peter Sollich and Srikanth Sastry},
      eprint={2505.14912},
      archivePrefix={arXiv},
      primaryClass={cond-mat.stat-mech},
      url={https://arxiv.org/abs/2505.14912}, 
      year={2025}
}

@article{BhaumikPNAS2021,
  title = {The role of annealing in determining the yielding behavior of glasses under cyclic shear deformation},
  author = {Himangsu Bhaumik and Giuseppe Foffi and Srikanth Sastry},
  journal = {PNAS},
  volume = {118},
  issue = {16},
  pages = {e2100227118},
  year = {2021},
  month = {April},
  doi = {https://doi.org/10.1073/pnas.210022711},
}

@article{ParmarPRX2019,
  title = {Strain Localization Above the Yielding Point in Cyclically Deformed Glasses},
  author = {Parmar, Anshul D. S. and Kumar, Saurabh and Sastry, Srikanth},
  journal = {Phys. Rev. X},
  volume = {9},
  issue = {2},
  pages = {021018},
  numpages = {11},
  year = {2019},
  month = {Apr},
  publisher = {American Physical Society},
  doi = {10.1103/PhysRevX.9.021018},
  url = {https://link.aps.org/doi/10.1103/PhysRevX.9.021018}
}

@article{LiuJCP2022,
    author = {Liu, Chen and Ferrero, Ezequiel E. and Jagla, Eduardo A. and Martens, Kirsten and Rosso, Alberto and Talon, Laurent},
    title = "{The fate of shear-oscillated amorphous solids}",
    journal = {The Journal of Chemical Physics},
    volume = {156},
    number = {10},
    pages = {104902},
    year = {2022},
    month = {03},
    issn = {0021-9606},
    doi = {10.1063/5.0079460},
    url = {https://doi.org/10.1063/5.0079460},
}

@Article{LAMMPS,
  author = "A. P. Thompson and H. M. Aktulga and R. Berger and 
     D. S. Bolintineanu and W. M. Brown and P. S. Crozier and
     P. J. in 't Veld and A. Kohlmeyer and S. G. Moore and T. D. Nguyen and
     R. Shan and M. J. Stevens and J. Tranchida and C. Trott and S. J. Plimpton",
  title = "{LAMMPS} - a flexible simulation tool for
     particle-based materials modeling at the 
     atomic, meso, and continuum scales",
  journal = "Comp. Phys. Comm.",
  volume =  "271",
  pages =   "108171",
  year =    "2022",
  doi = "10.1016/j.cpc.2021.108171"
}



\end{document}